\documentclass[10pt,journal,compsoc]{IEEEtran}
\ifCLASSOPTIONcompsoc
  \usepackage[nocompress]{cite}
\else
  \usepackage{cite}
\fi
\usepackage{url}

\ifCLASSINFOpdf
   \usepackage[pdftex]{graphicx}
\else
\fi
\usepackage{amsmath}
\ifCLASSOPTIONcompsoc
 \usepackage[caption=false,font=footnotesize,labelfont=sf,textfont=sf]{subfig}
\else
 \usepackage[caption=false,font=footnotesize]{subfig}
\fi

\usepackage{amssymb}
\usepackage{enumerate}
\usepackage{subfiles}
\usepackage{xcolor}
\usepackage{arydshln}
\usepackage[flushleft]{threeparttable}
\usepackage[export]{adjustbox}
\usepackage{listings}
\definecolor{codegreen}{rgb}{0,0.6,0}
\definecolor{codegray}{rgb}{0.5,0.5,0.5}
\definecolor{codepurple}{rgb}{0.58,0,0.82}
\definecolor{backcolour}{rgb}{0.95,0.95,0.92}
\lstdefinestyle{mystyle}{
    backgroundcolor=\color{backcolour},   
    commentstyle=\color{codegreen},
    keywordstyle=\color{magenta},
    numberstyle=\tiny\color{codegray},
    stringstyle=\color{codepurple},
    basicstyle=\ttfamily\footnotesize,
    breakatwhitespace=false,         
    breaklines=true,                 
    captionpos=b,                    
    keepspaces=true,                 
    numbers=left,                    
    numbersep=5pt,                  
    showspaces=false,                
    showstringspaces=false,
    showtabs=false,                  
    tabsize=2
}
\usepackage{tikz}
\usetikzlibrary{positioning}
\usepackage[hidelinks]{hyperref}

\newcommand\ie{\textit{i.e.}}
\newcommand\eg{\textit{e.g.}}
\newcommand\etal{\textit{et al.}}

\newcommand{\zhihao}[1]{#1}

\def\lossvaenll{\mathcal{L}}
\def\lossvae{\mathcal{L}_\lambda}
\def\lossvaevr{\mathcal{L}_\text{VR}}
\def\kldiv{D_\text{KL}}
\def\pdata{p_\text{data}}
\def\LRA{\Leftrightarrow}

\def\qzi{ q_i }
\def\pzi{ p_i }
\def\ourblock{latent block}

\def\decoder{ f_\text{dec} }

\begin{document}
%
\title{QARV: Quantization-Aware ResNet VAE for Lossy Image Compression}
%
%
%
%

\author{
Zhihao~Duan,
Ming~Lu,
Jack~Ma,
Yuning~Huang,
Zhan~Ma,
and~Fengqing~Zhu
\IEEEcompsocitemizethanks{
\IEEEcompsocthanksitem
Z. Duan, J. Ma, Y. Huang, and F. Zhu are with the Elmore Family School of Electrical and
Computer Engineering, Purdue University, West Lafayette, Indiana 47907, U.S.A.
\protect\\
E-mail: \{duan90, ma699, huan1781, zhu0\}purdue.edu.
\IEEEcompsocthanksitem
M. Lu and Z. Ma are with the School of Electronic Science and Engineering, Nanjing University, Nanjing, Jiangsu 210093, China.
\protect\\
E-mail: \{minglu, mazhan\}@nju.edu.cn.
\IEEEcompsocthanksitem
Research reported in this publication was supported by the National Cancer Institute (grant number: R01CA277839). The content is solely the responsibility of the authors and does not necessarily represent the official views of the National Institutes of Health.
\IEEEcompsocthanksitem
{\it Z. Duan and M. Lu equally contributed to this work.}
}
}

%
%

\markboth{IEEE Transactions on Pattern Analysis and Machine Intelligence, 2023}%
{Shell \MakeLowercase{\textit{et al.}}: Bare Demo of IEEEtran.cls for Computer Society Journals}
%



\IEEEtitleabstractindextext{%

\begin{abstract}
This paper addresses the problem of lossy image compression, a fundamental problem in image processing and information theory that is involved in many real-world applications.
We start by reviewing the framework of variational autoencoders (VAEs), a powerful class of generative probabilistic models that has a deep connection to lossy compression.
Based on VAEs, we develop a new scheme for lossy image compression, which we name quantization-aware ResNet VAE (QARV).
Our method incorporates a hierarchical VAE architecture integrated with test-time quantization and quantization-aware training, without which efficient entropy coding would not be possible.
In addition, we design the neural network architecture of QARV specifically for fast decoding and propose an adaptive normalization operation for variable-rate compression.
Extensive experiments are conducted, and results show that QARV achieves variable-rate compression, high-speed decoding, and better rate-distortion performance than existing baseline methods.
The code of our method is publicly accessible at \url{https://gitlab.com/viper-purdue/qarv-release}.
\end{abstract}

\vspace{-5pt}

}

\maketitle

\IEEEdisplaynontitleabstractindextext

%
\IEEEpeerreviewmaketitle


\IEEEraisesectionheading{\section{Introduction} \label{sec:qarv_introduction}}

%
%
%


\IEEEPARstart{E}{fficient} data storage/communication and generative modeling of data distributions are fundamentally connected through the principle of information compression.
Intuitively, the essence of data compression is to represent frequently-appeared symbols with short codewords and infrequent symbols with long ones.
To achieve good compression, one would need an accurate probabilistic model of the data distribution.
This objective of modeling data distribution coincides with likelihood-based generative models, a class of methods that play an important role in modern machine learning.
For example, recent studies have shown that many powerful generative models, such as auto-regressive models~\cite{zhang2021nelloc}, Variational Autoencoders~\cite{townsend2020hilloc}, Normalizing Flows~\cite{chang2021iflow}, and diffusion probabilistic models~\cite{kingma2021vdm}, can also serve as highly effective lossless compressors.


A similar conclusion holds for lossy compression as well.
In particular, Variational Autoencoders (VAEs) have been proven to employ a rate-distortion interpretation~\cite{alemi2018fixing_elbo} and have been widely adopted in recent, learning-based lossy compression methods~\cite{balle2016end2end, balle18hyperprior}.
However, most of the existing methods only retain up to a few (two or three) VAE layers and require designing a separate context model to improve compression performance.
Extensive research has been devoted to designing such context models, such as spatial auto-regressive models~\cite{minnen2018joint, he2021checkerboard} and channel-wise ones~\cite{minnen2020channelwise}, which not only are complex in design but also could be computationally intractable in practice~\cite{minnen2018joint}.

In this paper, we propose a hierarchical VAE-based framework for lossy image compression to address the issues discussed above.
Specifically, it avoids the use of spatial/channel-wise auto-regressive context models, attains high computational efficiency, yet still achieves better compression performance than existing methods.
The hierarchical design of our method is motivated by the fact that hierarchical VAEs (specifically, ResNet VAEs~\cite{kingma2016iafvae}) generalize autoregressive models and outperform them in distribution modeling~\cite{child2021vdvae}.
Hence, it is natural to expect them to succeed in lossy compression as well.
Adopting ResNet VAEs for lossy compression, however, is not a straightforward task, as their latent variables are continuously valued, and thus existing entropy coding algorithms cannot be applied.
Despite prior efforts to adopt ResNet VAEs to lossy compression, \eg, through Relative Entropy Coding~\cite{flamich2020rec} and post-training quantization~\cite{yang2020quantization}, they suffer different drawbacks and are sub-optimal in compression performance compared to the lossy compressors that utilize context models.

\zhihao{
In order to effectively leverage ResNet VAES for lossy compression, we design its probabilistic models and neural network architecture from the following three perspectives. 
First, entropy coding of latent variables is enabled by combining test-time quantization and quantization-aware training.
Specifically, the latent variables are discretized by elementwise uniform quantization at test time, and we simulate the quantization noise by additive uniform noise during training.
The VAE priors are parameterized using a Gaussian distribution convolved with a uniform distribution to accommodate the posteriors.
Under this configuration, we are able to turn ResNet VAEs into lossy image codecs while keeping all its advantages, \eg, its hierarchical architecture, progressive coding, and efficient computation, as demonstrated in our previous paper~\cite{duan2023qres}.
Second, we exploit the fact that ResNet VAEs naturally have a bias toward fast decoding (and slower encoding) due to their bi-directional inference structure, and we introduce a new block architecture to shift more computation from the decoder to the encoder.
This architecture allows our model to achieve a faster decoding speed than most previous image compression methods, even when ours is scaled to a deep hierarchy of latent variables.
Finally, we introduce a new variable-rate compression method, which we term adaptive layer normalization (AdaLN), that can be used in modern neural network architectures in a plug-and-play fashion.

Combining these features, we present Quantization-Aware ResNet VAE (QARV), a VAE model that achieves efficient lossy image compression at continuously adjustable rates using a single model.
QARV is conceptually similar to the Hyperprior model~\cite{balle18hyperprior} commonly used in learned image compression as both employ a hierarchical architecture.
However, QARV is fundamentally different in important aspects such as residual coding and bi-directional inference (details in Sec.~\ref{sec:qarv_related_hierarchical}), which make it more flexible and powerful.
Experimental results confirm that QARV outperforms existing published lossy image codecs (most of which are based on the Hyperprior with context models) in terms of BD-rate~\cite{bjontegaard2001bdrate} while retaining competitive computational efficiency.
}


\zhihao{
To summarize, our contributions are as follows (the first one is an extended analysis of our prior work~\cite{duan2023qres}, and the remaining ones are from this paper):
\begin{itemize}
    \item We give a comprehensive theoretical explanation of lossy compression using ResNet VAEs;
    \item We present a model architecture that achieves faster decoding and better R-D than existing baselines;
    \item We introduce a new method, adaptive layer normalization (AdaLN), for variable-rate compression;
    \item Overall, we propose a neural network model, namely QARV, that is simpler in design (no context models), more flexible (rate-variable, hierarchical features), better in BD-rate, and more efficient (fast CPU decoding) than most existing methods including the fixed-rate ones.
\end{itemize}
}

\section{Background: From Transform Coding to Coding with Generative Models} 
\label{sec:qarv_related}

Starting with traditional transform coding-based methods, we systematically review existing methods for lossy image compression.
We then summarize the connection between VAEs and lossy compression, which largely motivates the design of our approach.

\subsection{Lossy Image Compression by Transform Coding}
\label{sec:qarv_related_lic}
In conventional image codecs such as the JPEG~\cite{wallace1992jpeg}, JPEG 2000~\cite{Skodras2001jpeg2000}, and VVC intra~\cite{pfaff2021vvc_intra}, an orthogonal linear transformation such as discrete cosine transform or discrete wavelet transform is applied to convert image pixels into transform coefficients.
The transform coefficients are subsequently quantized and (losslessly) coded into bits using entropy coding algorithms.
To decode these bits, inverse operations (\eg, entropy decoding and the inverse transform) are applied to obtain a reconstruction.

Most learning-based image compression methods follow the same transform coding paradigm but in a non-linear way~\cite{balle2021nonlinear}.
In these methods, both the encoder and decoder are parameterized by deep neural networks (DNNs) instead of linear orthogonal transforms.
To enable rate-distortion optimization, an \textit{entropy model}, which is also parametrized by DNNs, is built for the compressed representation to estimate the number of bits needed for coding in a differentiable way.
To improve compression efficiency, existing works have explored efficient neural network blocks, such as residual networks~\cite{cheng2020cvpr}, self-attention~\cite{chen2021nlaic}, and transformers~\cite{lu2022tic, zou2022stf, lu2022tinylic}, as well as developed more expressive entropy models, such as hierarchical~\cite{balle18hyperprior, hu2022tpami_benchmark} and autoregressive models~\cite{minnen2018joint, minnen2020channelwise, he2021checkerboard, ma2022tpami_iwave}.
Interestingly, despite the fact that these methods are developed from the perspective of non-linear transform coding, their training framework resembles a simple (either one-layer~\cite{balle2016end2end} or two-layer~\cite{balle18hyperprior}) variational autoencoder (VAE), which we turn to next.
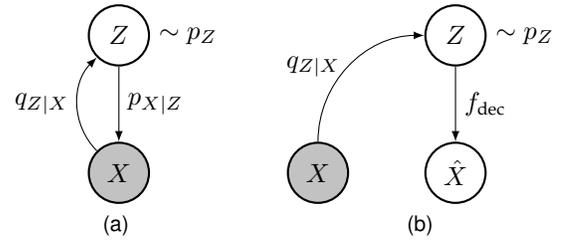
\begin{figure}[t]
\centering
    \hspace{+16pt}
    \subfloat[\label{fig:qarv_related_vae_original}]{
        \begin{tikzpicture}[
        rv_node/.style={circle, draw=black!100, thick, minimum size=8mm},
        ob_node/.style={circle, draw=black!100, thick, minimum size=8mm, fill=black!24},
        ]
        \node[ob_node] (x) {$X$};
        \node[rv_node] (z) [above=of x, label=right:$\sim p_Z$] {$Z$};
        \draw[-latex] (x.north west) to [out=135,in=225] node[left=0mm] {$q_{Z|X}$} (z.south west);
        \draw[-latex] (z.south) -- (x.north) node [midway, right] (emission) {$p_{X|Z}$};
        \end{tikzpicture}
    }
    \hfill
    \subfloat[\label{fig:qarv_related_vae_lossy}]{
        \begin{tikzpicture}[
        dt_node/.style={circle, draw=black!0, thick, minimum size=8mm},
        rv_node/.style={circle, draw=black!100, thick, minimum size=8mm},
        ob_node/.style={circle, draw=black!100, thick, minimum size=8mm, fill=black!24},
        ]
        \node[ob_node] (x)  {$X$};
        \node[rv_node] (r) [right=of x] {$\hat{X}$};
        \node[rv_node] (z)  [above=of r, label=right:$\sim p_Z$] {$Z$};
        \draw[-latex] (x.north) to [out=90,in=180] node[left=0mm] {$q_{Z|X}$} (z.west);
        \draw[-latex] (z.south) -- (r.north) node [midway, right] (emission) {$\decoder$};
        \end{tikzpicture}
    }
    \hspace{+16pt}
    \vspace{-0.08cm}
\caption{Probabilistic model of (a) a one-layer VAE in its original form and (b) a one-layer VAE for lossy compression. $X$ denotes data, $Z$ is the assumed latent variable, and $\hat{X}$ is the reconstruction.
}
\label{fig:qarv_related_vae}
\vspace{-0.24cm}
\end{figure}

\subsection{Variational Autoencoders (VAEs)}
\label{sec:qarv_related_vae}
VAEs~\cite{kingma14vae} are a class of latent variable models where conditional probability distributions are parameterized by neural networks.
Let $X$ be a random variable representing data (in our case, images) with an unknown data distribution $\pdata$.
In its original form (Fig.~\ref{fig:qarv_related_vae_original}), a VAE models data by a joint distribution $p_{X, Z}$:
\begin{equation}
    p_{X, Z}(x, z) = p_{X|Z}(x \mid z) \cdot p_Z(z),
\end{equation}
where $Z$ is an assumed \textit{latent variable}, $p_Z$ is the \textit{prior} distribution of $Z$, and $p_{X|Z}$ is the conditional data likelihood of $X$ given $Z$.
VAEs also define an \textit{approximate posterior}, $q_{Z|X}$, to perform variational inference of $Z$ given $X$.
The combination of $q_{Z|X}$ and $p_{X|Z}$ are also commonly interpreted as a (stochastic) autoencoder, where $q_{Z|X}$ is the encoder that encodes data $X$ into a latent variable $Z$, and $p_{X|Z}$ is the decoder that takes $Z$ and reconstructs $X$.

The goal of VAE is to find the set of model parameters such that $p_X$, the marginal distribution of $X$ under $p_{X, Z}$, is close to the true data distribution $\pdata$.
This is done by minimizing a tight upper bound~\cite{kingma14vae} on the negative log-likelihood (NLL) of $p_X$:
\begin{equation}
\label{eq:qarv_vae_loss}
\begin{aligned}
    \lossvaenll
    &\triangleq
    \mathbb{E}_{X \sim \pdata, Z \sim q_{Z|X}} \left[ 
    \kldiv(q_{Z|X} || \, p_Z) + \log \frac{1}{p_{X|Z}(X | Z)}
    \right]
    \\
    &\ge
    \mathbb{E}_{X \sim \pdata} \left[ -\log p_X(X) \right],
\end{aligned}
\end{equation}
where $\kldiv$ denotes the Kullback–Leibler (KL) divergence.
By convention, natural logarithms are used in Eq.~\eqref{eq:qarv_vae_loss},
so both NLL and $\lossvaenll$ have a unit of nats.
Note that the NLL term is the cross-entropy from $p_X$ to $\pdata$, \ie, the lossless compression rate of (a discrete) $X$ under the model $p_X$, which builds a connection between VAEs and lossless compression.
In fact, any VAE can function as a lossless compressor by using the bits-back coding algorithm~\cite{hinton1993bitsback}.

A different parameterization of VAE is used for lossy compression, which is shown in Fig.~\ref{fig:qarv_related_vae_lossy}.
Instead of $p_{X|Z}$, a deterministic decoder $\decoder$ and a lossy reconstruction $\hat{X} = \decoder(Z)$ are introduced. Under this formulation, the loss function becomes
\begin{equation}
\label{eq:qarv_vae_rd_loss}
    \lossvae =
    \mathbb{E}_{X \sim \pdata, Z \sim q_{Z|X}} \left[ 
    \kldiv(q_{Z|X} || \, p_Z)
    +
    \lambda \cdot d(X,\hat{X})
    \right],
\end{equation}
where $d(X,\hat{X})$ is a distortion metric (\eg, mean squared error) and $\lambda$ is a positive scalar.
Note that $\lossvae$ and $\lossvaenll$ are equivalent up to a constant additive factor, by simply setting $p_{X|Z} \propto \exp{ (-\lambda \cdot d(X,\hat{X}) ) }$ in Eq.~\eqref{eq:qarv_vae_loss}.

Recent works~\cite{alemi2018fixing_elbo, yang2022sandwich} have proved that Eq.~\eqref{eq:qarv_vae_rd_loss} is a tight upper bound on the Lagrangian relaxation of the information rate-distortion function~\cite{cover2005information}, showing the strong connection between the VAE framework and lossy compression.
This perhaps explains the success of VAEs in learned image compression, and it motivates us to use a more powerful VAE architecture---the deep hierarchical VAEs---for lossy image compression.

\subsection{Hierarchical VAEs} \label{sec:qarv_related_hierarchical}
To model high-dimensional data such as images, hierarchical VAEs~\cite{Sonderby2016laddervae, kingma2016iafvae, vahdat2020nvae, child2021vdvae} have been proposed to improve the flexibility and expressiveness of VAEs.
A hierarchical VAE employs a group of latent variables, denoted by $Z_{1:N} \triangleq \{Z_1, Z_2, ..., Z_N\}$, where $N$ is the total number of variables in an autoregressive manner:
\begin{equation}
    p_{Z_{1:N}} = p_{Z_N|Z_{<N}} \, \cdots \ p_{Z_3|Z_2,Z_1} \cdot p_{Z_2|Z_1} \cdot p_{Z_1},
\end{equation}
where $Z_{<N}$ denotes $\{Z_1, Z_2, ..., Z_{N-1}\}$.
Typically, $Z_1$ has a small number of dimensions, and $Z_N$ has a larger number of dimensions.
This low- to high-dimensional architecture not only improves the flexibility of VAEs but also captures the coarse-to-fine nature of images.

\begin{figure}[t]
    \centering
    \hfill
    \subfloat[]{
        \includegraphics[height=4.5cm]{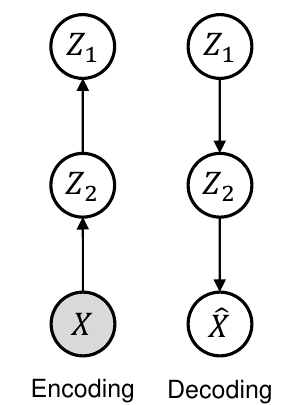}
        \label{fig:qarv_related_hyp}
    }
    \hfill
    \subfloat[]{
        \includegraphics[height=4.5cm]{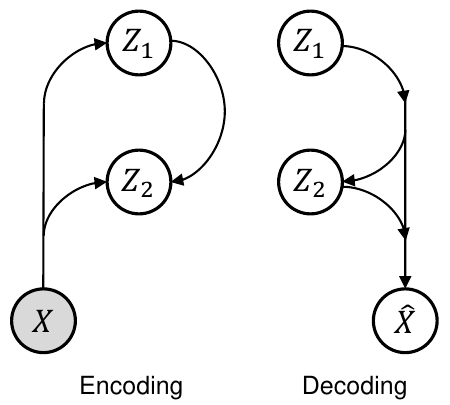}
        \label{fig:qarv_related_res}
    }
    \hfill
    \caption{Illustration of (a) a two-layer hyperprior VAE, and (b) a two-layer ResNet VAE. Detailed architecture of our ResNet VAE model is introduced in Sec.~\ref{sec:qarv_method}.}
    \label{fig:qarv_related_hierarchical}
\end{figure}

ResNet VAE~\cite{kingma2016iafvae}, a class of hierarchical VAEs, provides the most promising performance in terms of NLL on images~\cite{child2021vdvae, sinha2021crvae}.
Fig.~\ref{fig:qarv_related_res} shows the encoding and decoding processes of a ResNet VAE for lossy compression. As a comparison, we also show the Hyperprior VAE model~\cite{balle18hyperprior} in Fig.\ref{fig:qarv_related_hyp}.
ResNet VAEs are fundamentally different from Hyperprior VAEs in several important aspects:
\vspace{-1mm}
\begin{itemize}
    \item Both the encoding and decoding use the same top-down ordering of latent variables (while in Hyperprior VAE the two orderings are reversed);
    \item Each latent variable $Z_i$ is conditionally dependent on all $Z_{<i}$ (while in Hyperprior VAE the dependency is first-order Markov);
    \item The encoding is bi-directional, \ie, $Z_i$ is dependent on both $X$ and $Z_{<i}$ (while in Hyperprior VAE the encoding is bottom-up).
\end{itemize}
\vspace{-1mm}
These differences might account for the posterior collapse problem in Hyperprior-style models~\cite{Sonderby2016laddervae}, resulting in the fact that ResNet VAEs can be scaled up to more than 70 layers~\cite{child2021vdvae}, while only three layers have been done for the Hyperprior model~\cite{hu2022tpami_benchmark}.

The loss function for training ResNet VAEs is the same as Eq.~\eqref{eq:qarv_vae_rd_loss} but extended for multiple latent variables:
\begin{equation}
\label{eq:qarv_related_resnet_vae_loss}
    \lossvae =
    \mathbb{E}_{X \sim \pdata, Z_{1:N} \sim q_{1:N}} \left[ 
    \sum_{i=1}^N \kldiv(q_i || \, p_i)
    +
    \lambda \cdot d(X,\hat{X})
    \right],
\end{equation}
where $\qzi$ and $\pzi$ are shorthand notations for the posterior and prior for the $i$-th latent variable, \ie,
\begin{equation}
\label{eq:qarv_related_notation}
\begin{aligned}
    \qzi (\cdot) & \triangleq q_{Z_i|X,Z_{<i}} (\cdot)
    \\
    \pzi (\cdot) & \triangleq p_{Z_i|Z_{<i}} (\cdot).
\end{aligned}
\end{equation}
In practice, the expectation w.r.t. $Z_{1:N}$ in Eq.~\eqref{eq:qarv_related_resnet_vae_loss} is estimated by drawing samples of $z_{1:N}$ from $q_{1:N}$. This is implemented by ancestral sampling, \ie, we iteratively sample $z_i$ from $q_i$ for $i = 1, 2, ..., N$, where at each step $q_i$ is conditional on the previously sampled value of $z_{<i}$.


\subsection{Practical Coding with VAEs}
\label{sec:qarv_related_vae_coding}
Using the bits-back coding algorithm~\cite{hinton1993bitsback}, VAEs can readily perform lossless compression of discrete $X$ with a rate equal to $\lossvae$.
Existing methods have successfully applied VAEs, including a single-layer VAE~\cite{townsend2018bbans} and hierarchical ones~\cite{kingma2019bitswap, townsend2020hilloc}, to lossless compression of images.
For lossy compression, however, bits-back coding is not applicable, and lossy compression using (hierarchical) VAEs remains a difficult problem to solve.

There are several lines of work that try to apply VAEs to lossy data compression.
One approach is to implement relative entropy coding (REC)~\cite{flamich2020rec}, a coding scheme where one wants to code stochastic samples of VAE latent variables with an average codelength close to the KL term in Eq.~\eqref{eq:qarv_related_resnet_vae_loss}.
Extensive research~\cite{agustsson2020universally, flamich2020rec, theis2022algorithms_comm_sapmles, flamich2022a_star_rec} has been conducted towards this goal, but they either require an intractable execution time or incur a large codelength overhead.
Another direction is to develop VAEs with discrete latent variables.
Discrete VAEs (DVAEs)~\cite{rolfe2016dvae, vahdat2018dvaepp, vahdat2018dvaehash} assume Bernoulli latent variables, and Vector Quantized VAEs (VQ-VAEs)~\cite{oord2017vqvae, razavi2019vqvae2, williams2020hqa} assume categorical latent variables.
Due to various statistical challenges, however, DVAEs and VQ-VAEs are still not capable of end-to-end R-D optimization in their current form.

As we mentioned in Sec.~\ref{sec:qarv_related_lic}, many learned image codecs could also be interpreted as a (either one-layer or two-layer) Hyperprior VAE~\cite{balle18hyperprior}.
In these methods, the latent variables are continuous during training but quantized during testing to enable entropy coding.
Inspired by this, we developed a similar coding strategy to turn ResNet VAEs into practical lossy image compressors, which brings several advantages to learned image compression, such as multi-layer progressive coding and efficient execution.

\subsection{Variable-Rate Learned Image Compression}

Various methods have been proposed to realize variable-rate compression using a single model.
In an early work, Choi \etal{} introduce a conditional autoencoder with adjustable quantization bin size~\cite{choi2019var_rate_conditional_ae}.
Chen and Ma propose to apply a post-training affine transform to modulate compressed coefficients~\cite{chen2020var_rate_quality_scaling}.
Yang \etal{} propose a slimmable architecture with adjustable network complexity and compression rate~\cite{yang2021slimmable}.
Song \etal{} use a quality map to condition the compression model, enabling spatially-adaptive rate adjustment~\cite{song2021qmap}.
More recently, Cai \etal{} introduce an invertible neural network module that is conditioned on the quality level~\cite{cai2022var_rate_invertible_activation}.
Lee \etal{} use a learnable mask to select the subset of compressed coefficients to compress~\cite{lee2022selective}.
Gao \etal{} directly optimize the compressed coefficients by semi-amortized inference for each specific R-D objective~\cite{gao2022flexible}.
In this paper, we introduce a new approach to variable rate compression
(Sec.~\ref{sec:qarv_method2_rate_conditional}), which is simple yet effective.

\begin{figure*}[ht]
    \centering
    \subfloat[Training]{
        \includegraphics[width=0.32\linewidth]{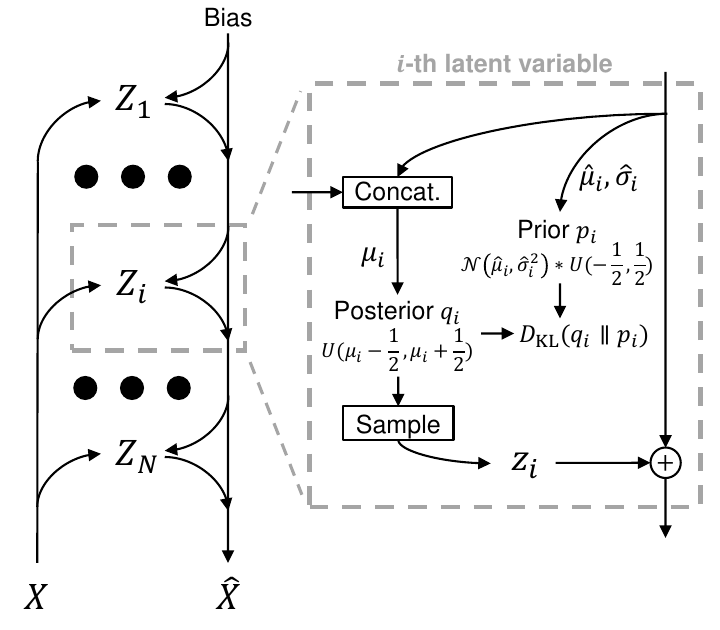}
        \label{fig:qarv_method_overview_training}
    }
    \hfill
    \subfloat[Testing: compress]{
        \includegraphics[width=0.32\linewidth]{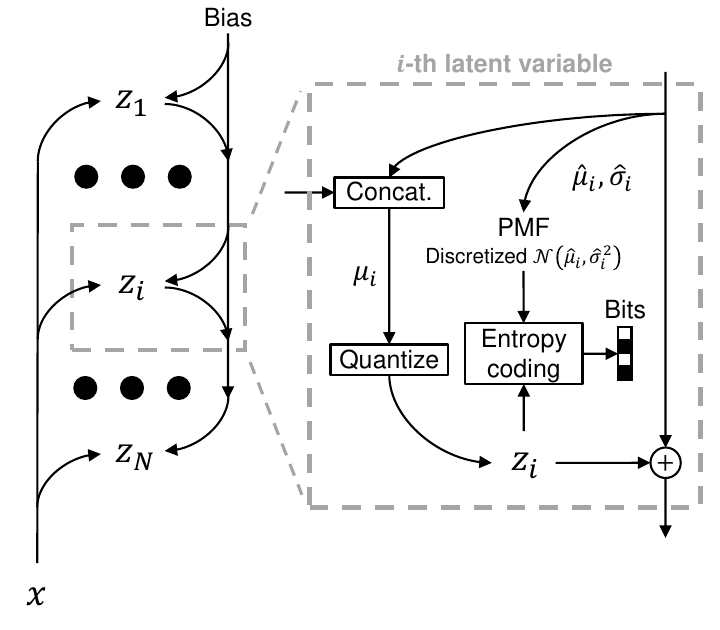}
        \label{fig:qarv_method_overview_compress}
    }
    \hfill
    \subfloat[Testing: decompress]{
        \includegraphics[width=0.32\linewidth]{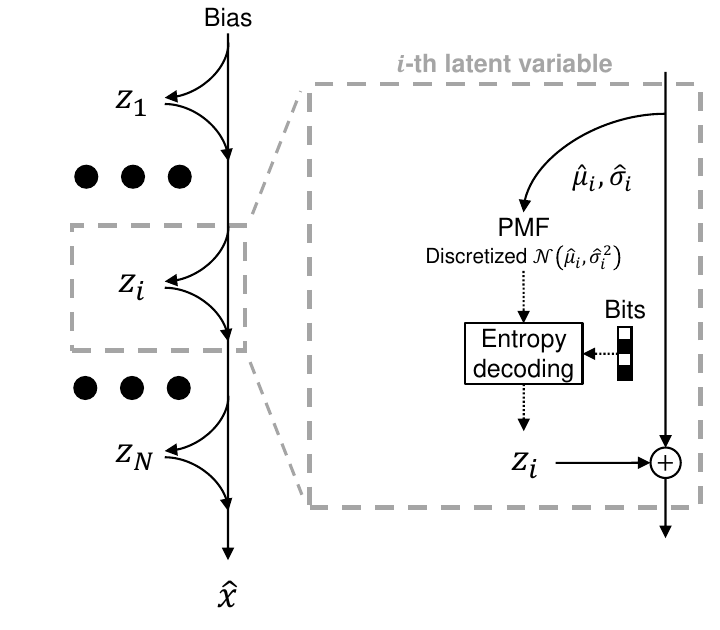}
        \label{fig:qarv_method_overview_decompress}
    }
    \caption{\textbf{Illustration of the QARV framework.} QARV has thee modes: (a) training, (b) compressing, and (c) decompressing.
    During training, QARV learns to minimize the KL divergence between all pairs of priors and posteriors, as well as the distortion between $\hat{X}$ and $X$.
    For practical compression, the uniform posteriors are replaced by uniform quantization, and the priors are turned into discrete PMFs for entropy coding.
    }
    \label{fig:qarv_method_overview}
\end{figure*}

\section{Quantization-Aware ResNet VAE (QARV)}
\label{sec:qarv_method}

As we described earlier, VAEs with continuous latent variables cannot be readily used for lossy compression since all existing entropy coding algorithms, \eg, Huffman coding~\cite{huffman1952}, require a discrete alphabet.
As an overview, we combat this issue by applying simple quantization (elementwise uniform) at test time, which produces a discrete alphabet and thus enables entropy coding. To make training quantization aware, we use uniform noise to model the quantization errors, which is a standard technique in learned image compression~\cite{balle2016end2end}.

We start with presenting the probabilistic model and loss function used during training in Sec.~\ref{sec:qarv_method_probabilistic}.
We describe how practical entropy coding is done in Sec.~\ref{sec:qarv_method_compression}.
Finally, we show that QARV generalizes commonly-used context models for learned image compression in Sec.~\ref{sec:qarv_method_compression}.


\subsection{A Quantization-Aware Probabilistic Model} \label{sec:qarv_method_probabilistic}
The probabilistic model for training QARV is depicted in Fig.~\ref{fig:qarv_method_overview_training}.
It employs a hierarchy of $N$ latent variables, denoted by $Z_1, Z_2, ..., Z_N$.
In the encoding process, a sequence of posteriors $q_1, q_2, ..., q_N$ are applied to infer $Z_1, Z_2, ..., Z_N$ autoregressively given the data $X$.
For every latent variable, a (conditional) uniform posterior is used to model the uniform quantization error that will be encountered during testing. We now give the formal definitions below.

\textbf{Posteriors:}
The posterior distribution of $Z_i$ given $x$ and $z_{<i}$ is defined to be a uniform distribution:
\begin{equation}
\label{eq:qarv_method_posterior}
\begin{aligned}
\qzi &\triangleq U(\mu_i - \frac{1}{2}, \mu_i + \frac{1}{2})
\\
\LRA \qzi(z_i \mid z_{<i},x) &= 
\left\{\begin{matrix}
1 & \text{for} \ \left | z_i - \mu_i \right | \leq \frac{1}{2} \\ 
0 & \text{otherwise}
\end{matrix}\right. ,
\end{aligned}
\end{equation}
where the parameter $\mu_i$ is the output from the posterior branch in the $i$-th \ourblock{} in Fig.~\ref{fig:qarv_method_overview_training}.
Note that the dependency of $\qzi$ on $x$ and $z_{<i}$ is made through $\mu_i$, which is a function of both $x$ and $z_{<i}$, as we can see in Fig.~\ref{fig:qarv_method_overview_training}.
Once $\qzi$ is obtained, a $z_i$ is sampled and aggregated to the top-down path through an addition operation.
The sampling can be easily implemented via additive uniform noise:
\begin{equation}
\label{eq:qarv_method_posterior_sampling}
    z_i \leftarrow \mu_i + u ,
\end{equation}
where $u$ is a random sample from $U(-\frac{1}{2}, \frac{1}{2})$.
This process of 1) computing the posterior $\qzi$, 2) sampling $z_i$ from it, and 3) aggregating $z_i$ into the top-down path is done iteratively for all latent variables in sequential order, which is also known as the \textit{ancestral sampling} used in autoregressive models.

In addition to the posteriors, the model produces a prior distribution, $\pzi$, for every latent variable $Z_i$.
As the VAE loss function (Eq.~\eqref{eq:qarv_related_resnet_vae_loss}) suggests, a KL divergence term between $\qzi$ and $\pzi$ for all $i \in \{1, 2, ..., N\}$ is used during training, which imposes several requirements for choosing a proper parametric form of $\pzi$.
First, the probability density function (pdf) of $\pzi$ should be flexible enough to resemble a uniform distribution (since we want to minimize $\kldiv(\qzi \parallel \pzi)$).
Second, the pdf must also be non-zero everywhere; otherwise, evaluating $\kldiv(\qzi \parallel \pzi)$ may result in a division by zero, causing numerical instabilities in training.
Third, it should be easy to turn $\pzi$ into a discrete probability mass function (pmf), which will be used for entropy coding the quantized $Z_i$ during testing.
A good choice that satisfies these requirements is given by Ball\'{e} \etal{} in~\cite{balle18hyperprior, minnen2018joint}.
We adopt a similar construction and extend it to the ResNet VAE architecture, which we give as follows.

\textbf{Priors:}
The prior distribution $\pzi$ is chosen to be a (conditional) Gaussian convolved with a uniform distribution:
\begin{equation}
\label{eq:method_prior}
\begin{aligned}
\pzi &\triangleq \mathcal{N}(\hat{\mu}_i, \hat{\sigma}_i^2) * U(- \frac{1}{2}, \frac{1}{2})
\\
\LRA \pzi(z_i \mid z_{<i}) &= \int_{z_i-\frac{1}{2}}^{z_i+\frac{1}{2}} \mathcal{N}(t; \hat{\mu}_i, \hat{\sigma}_i^2) \ dt ,
\end{aligned}
\end{equation}
where $\mathcal{N}(t; \hat{\mu}_i, \hat{\sigma}_i^2)$ denotes the Gaussian pdf with mean $\hat{\mu}_i$ and standard deviation $\hat{\sigma}_i$ evaluated at $t$, {and $t$ is an integration dummy variable}.
The conditional dependency of $\pzi$ on $z_{<i}$ is made through $\hat{\mu}_i$ and $\hat{\sigma}_i$, which are the outputs of the prior branch in the right pane of Fig.~\ref{fig:qarv_method_overview_training}.

To see how this $\pzi$ satisfies the three requirements we mentioned earlier, first notice that when $\hat{\mu}_i \rightarrow 0$ and $\hat{\sigma}_i \rightarrow 0$, the Normal pdf converges to a Dirac delta function, and thus $\pzi$ converges to $U(- \frac{1}{2}, \frac{1}{2})$, satisfying our first requirement.
It is also easy to see that $\pzi$ is positive everywhere (since it is integral over a positive function), which satisfies our second requirement.
Finally, $\pzi$ can be easily turned into a pmf, denoted by $P_i$, which is defined as
\begin{equation}
P_i(n) \triangleq p_{i}(\hat{\mu}_i + n| z_{<i}), n \in \mathbb{Z}.
\end{equation}
In other words, $P_i$ is obtained by evaluating $\pzi$ at evenly spaced values with unit step sizes.
Note that $P_i$ is a valid pmf, or specifically, a discretized Gaussian~\cite{balle18hyperprior}.
The pmf $P_i$ is used for entropy coding of $z_i$, which we describe in the next subsection.



\textbf{Decoder:} With $Z_{1:N}$ at hand, the reconstruction $\hat{X}$ is obtained by applying a (deterministic) decoder:
\begin{equation}
    \hat{X} = \decoder(Z_{1:N}),
\end{equation}
where $\decoder$ corresponds to the top-down path (from the Bias to $\hat{X}$) in Fig.~\ref{fig:qarv_method_overview_training}.
Notice that the top-down path is shared between encoding and decoding, which leads to the fact that decoding is much faster than encoding (because encoding contains most part of the decoding). We exploit this feature in our architecture design in Sec.~\ref{sec:qarv_method2_block_fast_dec} to further shift computation from decoding to encoding.

\zhihao{
\textbf{Training objective (fixed-rate).}
The training objective is to minimize the ResNet VAE loss function $\lossvae$ in Eq.~\eqref{eq:qarv_related_resnet_vae_loss}.
Using the posteriors and priors defined before, $\lossvae$ for QARV can be expressed as
\begin{align}
    \lossvae
    &= \mathbb{E}_{X, Z_{1:N}} \left[ 
    \sum_{i=1}^N \kldiv(q_i || \, p_i) + \lambda \cdot d(X,\hat{X})
    \right]
    \\
    &= \mathbb{E}_{X, Z_{1:N}} \left[
    \sum_{i=1}^N \log \frac{\qzi(Z_i \mid Z_{<i},X)}{\pzi(Z_i \mid Z_{<i})} + \lambda \cdot d(X, \hat{X})
    \right]
    \\
    &= \mathbb{E}_{X, Z_{1:N}} \left[
    \sum_{i=1}^N \log \frac{1}{\pzi(Z_i \mid Z_{<i})}
    +
    \lambda \cdot d(X, \hat{X})
    \right],
    \label{eq:qarv_method_loss}
\end{align}
where the expectation is w.r.t. $X \sim \pdata$ and
\begin{equation}
    Z_{1:N} \sim q_N \cdots q_2 \cdot q_1.
\end{equation}
The first term in Eq.~\eqref{eq:qarv_method_loss} consists of the rate for all latent variables, each of which is a continuous relaxation of their test-time bit rate (up to a constant factor).
The second term corresponds to the reconstruction distortion, which is commonly chosen to be the mean squared error for images.
We refer to Eq.~\eqref{eq:qarv_method_loss} as the \textit{fixed-rate training objective} since the multiplier $\lambda$, which trades off rate and distortion, is pre-determined and fixed throughout training.
As a result, each separately trained model can only compress images at a specific rate.
To compress images at different rates, multiple models have to be trained, which is inflexible in real-world applications and motivates a variable-rate training objective.

}

\textbf{Training objective (variable-rate).}
We adopt the conditional VAE~\cite{sohn2015conditional_vae, choi2019var_rate_conditional_ae} formulation for variable-rate compression.
In this scheme, a single model is trained to operate on a range of rates by accepting $\lambda$ as an input to the model, and all posteriors and priors are conditional on $\lambda$.
The goal of training is to optimize the conditional posteriors, $\qzi (z_i | x,z_{<i}, \lambda)$, and conditional priors, $\pzi (z_i | z_{<i}, \lambda)$, for the rate-distortion objective $\lossvae$.
To support continuously variable-rate compression, we randomly sample $\lambda$ from a continuous value range $[\lambda_\text{low}, \lambda_\text{high}]$ throughout training.
Formally, the variable-rate loss function is:
\begin{equation}
\label{eq:qarv_method_loss_var_rate}
    \lossvaevr
    = \mathbb{E}_{X, \Lambda, Z_{1:N}} \left[
    \sum_{i=1}^N \log \frac{1}{\pzi(Z_i | Z_{<i}, \Lambda)} + \Lambda \cdot d(X, \hat{X})
    \right],
\end{equation}
where $p_\Lambda$ is a human-specified pdf defining the sampling strategy of $\lambda$ throughout training.
Once the model is trained, we are able to adjust the rate-distortion trade-off using a single model by simply varying the $\lambda$ input to the model.
{We discuss our choice of $p_\Lambda$ and how we make the priors and posteriors conditional on $\lambda$ later in Sec.~\ref{sec:qarv_method2_rate_conditional}.}

\subsection{Compression and Decompression}
\label{sec:qarv_method_compression}
The QARV model can be easily converted into a lossy compressor by quantizing and entropy coding the latent variables in a way similar to the Hyperprior model~\cite{balle18hyperprior}.
We discuss the fixed-rate probabilistic model in this subsection, but the same process generalizes to our variable-rate model by conditioning all posteriors and priors on $\lambda$.

Fig.~\ref{fig:qarv_method_overview_compress} shows the procedure of compressing a data sample $x$ into bits using QARV.
Instead of sampling from the posterior, we obtain $z_i$ by quantizing the residual of $\mu_i$ w.r.t. $\hat{\mu}_i$:
\begin{equation}
z_i \leftarrow \hat{\mu}_i + \lfloor \mu_i - \hat{\mu}_i \rceil,
\end{equation}
where $\lfloor \cdot \rceil$ denotes nearest integer rounding.
In other words, $\mu_i$ is quantized to its nearest neighbor from the set
$\{ \hat{\mu}_i + n \mid n \in \mathbb{Z} \}$.
Then, $z_i$ is encoded into bits using standard entropy coding algorithms, with the pmf $P_i$.
Since each latent variable produces a separate bitstream, a compressed image consists of $N$ bitstreams, corresponding to the $N$ latent variables $z_1, z_2, ..., z_N$.


The procedure of decompression is shown in Fig.~\ref{fig:qarv_method_overview_decompress}.
Starting with a constant bias (which is learned from training), we iteratively compute $P_i$ for $i = 1,2,..., N$.
At each step, we apply the entropy decoding algorithm with pmf $P_i$ to decode $z_i$ from the $i$-th bitstream.
Note that $P_i$ at the decoding side is the same as the one used in encoding, so exact $z_i$ can be recovered.
We obtain the reconstruction $\hat{x}$ using the decoder $\decoder$ once all $z_{1:N}$ are decoded.



\begin{figure}[t]
    \centering
    \subfloat[]{
        \includegraphics[height=5cm]{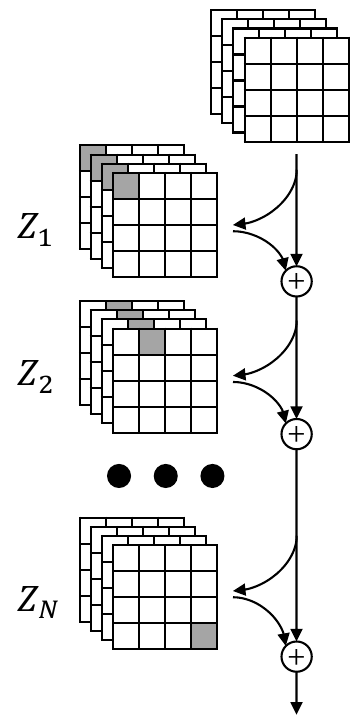}
        \label{fig:qarv_method_generalize_ar}
    }
    \hfill
    \subfloat[]{
        \includegraphics[height=5cm]{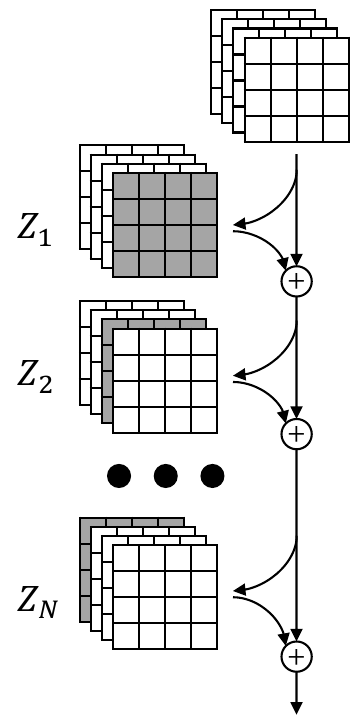}
    }
    \hfill
    \subfloat[]{
        \includegraphics[height=5cm]{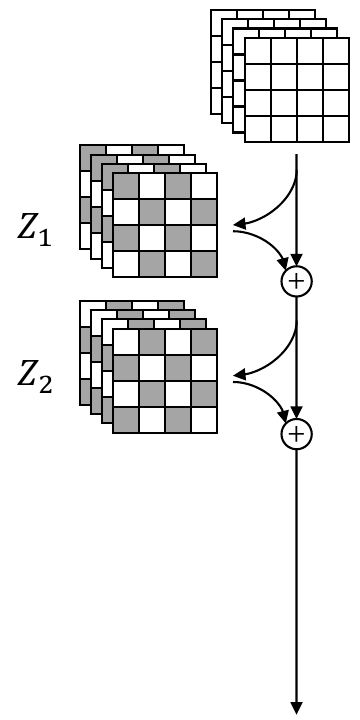}
    }
    \hfill
    \caption{Illustration of how QARV generalizes the AR context models in learned image compression methods \zhihao{given a fixed image resolution}, such as (a) the raster scan order AR model~\cite{minnen2018joint}, (b) the channel-wise AR model~\cite{minnen2020channelwise}, and (c) the Checkerboard model~\cite{he2021checkerboard}. White blocks represent zeros, and gray blocks represent non-zero values.}
    \label{fig:qarv_method_generalize}
\end{figure}

\subsection{QARV Generalizes AR Context Models}
\label{sec:qarv_method_generalize}
A nice property of ResNet VAEs is that, if sufficiently deep, it generalizes autoregressive (AR) models~\cite{child2021vdvae}.
As a result, one can also view QARV (for both the fixed-rate and variable-rate versions) as a generalization of the AR context models that are extensively used in existing learned image compression methods.
This conclusion holds for all types of autoregressive models, including the standard, raster scan order AR model~\cite{minnen2018joint} \zhihao{(for a constant image resolution)}, the channel-wise AR model~\cite{minnen2020channelwise}, and the Checkerboard context model~\cite{he2021checkerboard}.
We illustrate this fact in Fig.~\ref{fig:qarv_method_generalize}.
Taking Fig.~\ref{fig:qarv_method_generalize_ar} as an example.
\zhihao{
Assuming the feature vector has a fixed spatial dimension of $h\times w$.
By setting $N = h \cdot w$, QARV is feasible to predict the top-left pixel by $Z_1$, the pixel to its right by $Z_2$, and similar for $Z_3, Z_4, ..., Z_N$.
Adding $Z_{1:N}$ together, one obtains a complete feature vector, and this process resembles the spatial, raster scan order AR model~\cite{minnen2018joint}.
}
A similar conclusion also holds for the other two examples in Fig.~\ref{fig:qarv_method_generalize}, and we refer to~\cite{child2021vdvae} for a more rigorous proof.

In principle, ResNet VAEs (and thus QARV) are able to learn better latent representations with possibly much fewer layers than AR models~\cite{child2021vdvae}.
Thus, QARV has the potential to be more effective than existing AR-based methods, at the same time avoiding the manual process of designing the autoregressive order in context models.






\section{Neural Network Implementation}
\label{sec:qarv_method2}

\begin{figure*}[ht]
    \centering
    \hfill
    \subfloat[Architecture overview]{
        \includegraphics[width=0.9\linewidth]{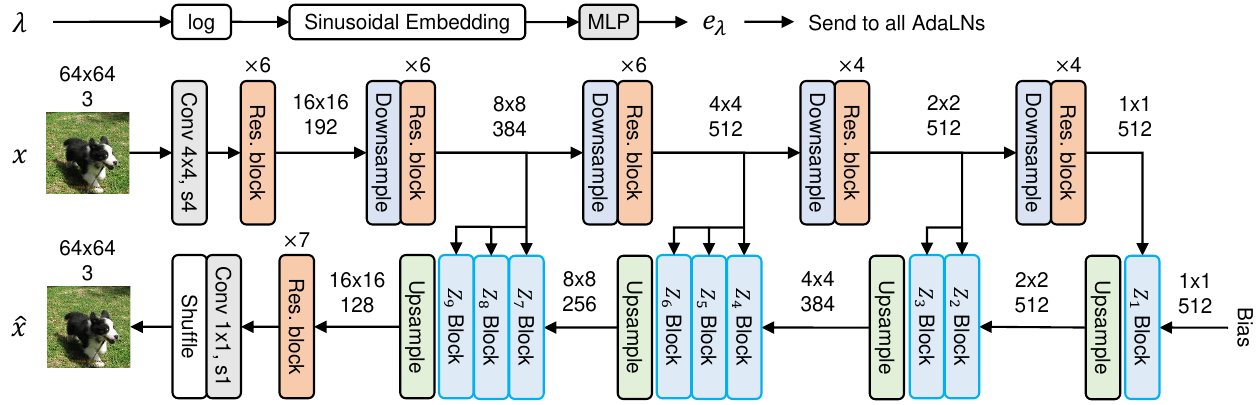}
        \label{fig:qarv_method_architecture_overall}
    }
    \hfill
    \vspace{-2mm}
    \\
    \hspace{1mm}
    \subfloat[Latent variable block (training)]{
        \includegraphics[height=54mm]{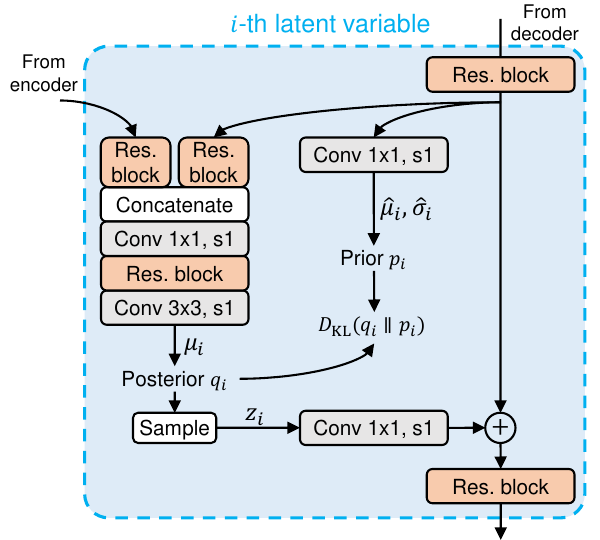}
        \label{fig:qarv_method_architecture_latent_block}
    }
    \hspace{8mm}
    \subfloat[Down/upsample]{
        \includegraphics[height=54mm]{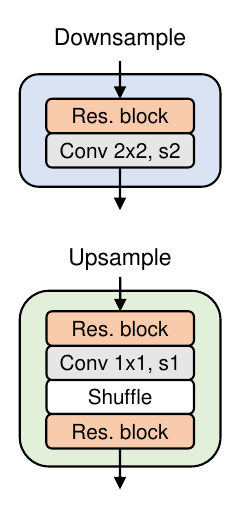}
        \label{fig:qarv_method_architecture_down_up}
    }
    \hspace{8mm}
    \subfloat[Res. block and AdaLN]{
        \includegraphics[height=54mm]{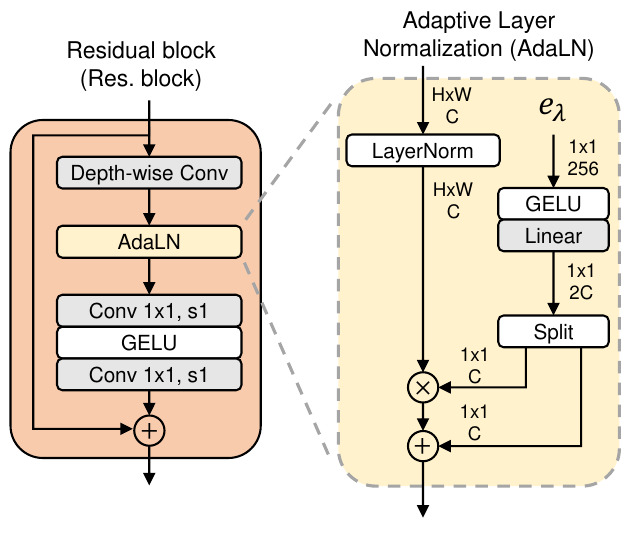}
        \label{fig:qarv_method_architecture_res_adaln}
    }
    \hspace{6mm}
    \caption{
    \zhihao{
    \textbf{Neural network architecture of QARV.} White blocks denote nonparametric operations. Colored (including gray) blocks denote learnable neural network modules. The spatial dimensions of features (height $\times$ width) scale accordingly with input image resolution, and in the figure, they are shown for a $64\times 64 \times 3$ (height, width, and channel) input.
    Detailed explanations are provided in Sec.~\ref{sec:qarv_method2}.
    }
    }
    \label{fig:qarv_method_architecture}
\end{figure*}

This section presents the neural network architecture we use to implement the (variable-rate) QARV model.
We focus on variable-rate compression, as fixed-rate models are both inefficient in training and impractical in real-world applications.
Our model architecture and its components are shown in Fig.~\ref{fig:qarv_method_architecture}, and we describe each of its components in detail in the following subsections.

\subsection{Overall Architecture}
Fig.~\ref{fig:qarv_method_architecture_overall} overviews the QARV model architecture.
The model accepts two inputs, an image $x$ to be compressed and a Lagrange multiplier $\lambda$ that trade-offs rate and distortion.
We overview the main model (lower part of Fig.~\ref{fig:qarv_method_architecture_overall}) in this subsection, and we introduce the $\lambda$ embedding network (top part of Fig.~\ref{fig:qarv_method_architecture_overall}) in Sec.~\ref{sec:qarv_method2_rate_conditional}.

The main model architecture follows the QARV framework described in Sec.~\ref{sec:qarv_method}.
On the encoding (sender) side, a CNN (left-to-right path in the figure) extracts a hierarchy of four features from the image, each having a different spatial resolution.
We chose the feature resolutions to be 8x, 16x, 32x, and 64x downsampled w.r.t. the original input image resolution.
For example, in Fig.~\ref{fig:qarv_method_architecture_overall}, where the input image has $64\times 64$ pixels, the feature resolution ranges from $8\times 8$ to $1 \times 1$.
These extracted features are then sent to the latent variable blocks in the decoding path (the right-to-left path) to produce the compressed bitstream.
Each latent variable block, indexed by $i$, contains the latent variable $Z_i$ as well as its posterior $\qzi$ and prior $\pzi$,
and each latent variable produces a separate sequence of bits during encoding. The collection of all such sequences forms the final bitstream of the input image $x$.

On the decoding (receiver) side, only the right-to-left path in Fig.~\ref{fig:qarv_method_architecture_overall} is needed to decode the bitstream. The latent variable blocks and upsampling layers are executed in their sequential order, and a reconstruction image $\hat{x}$ is obtained at the end of the path.
This architecture naturally offers a bias towards faster decoding than encoding since the entire network up to $Z_9$ is executed during encoding, while only the right-to-left path is executed during decoding.

Our network architecture is fully convolutional, so when the input image resolution is multiples of $64 \times 64$, all feature resolution scales accordingly.
For arbitrary resolution images, we simply pad the image (by replicating the edges) on the encoding side such that its resolution is multiples of $64 \times 64$. On the decoding side, we crop the reconstructed image to recover the original, un-padded image.

\subsection{Latent Variable Blocks: Shifting Computation from Priors to Posteriors}
\label{sec:qarv_method2_block_fast_dec}
In many practical applications, a high decoding speed is more desirable than encoding speed, and ResNet VAEs naturally offer such bias toward fast decoding (as can be seen from Fig.~\ref{fig:qarv_method_overview_compress} and Fig.~\ref{fig:qarv_method_overview_decompress}).
To further promote this fact, we design a latent variable block architecture specifically for efficient decoding by shifting computation from the priors to the posteriors.

The network structure of the priors and posteriors is shown in Fig.~\ref{fig:qarv_method_architecture_latent_block}.
The left-hand side of the figure shows the posterior branch for each latent variable, and the middle part of the figure shows the corresponding prior branch.
The posterior branch consists of three residual blocks, a concatenation operation, and two convolutional layers, a combination of which fuses the features from the encoding path and the decoding path.
In contrast, the prior branch contains only a single convolutional layer.
Recall that both the posterior branch and the prior branch need to be executed for encoding $z_i$, but only the prior branch is needed for decoding $z_i$.
Thus, by employing such a block architecture, our model realizes much more efficient decoding than encoding.
In the experiments (Sec.~\ref{fig:qarv_exp_main}), we show that our model decodes images at around 2x to 3x the speed of encoding, and the decoding speed is comparable to or outperforms previous learning-based methods on both CPUs and GPUs.

\subsection{Network Components}

The downsampling and upsampling operators are shown in Fig.~\ref{fig:qarv_method_architecture_down_up}.
For downsampling, we use a convolutional layer whose stride equals the kernel size.
This operator is also known as the patch embedding layer, which is commonly used in Vision Transformer architectures~\cite{dosovitskiy2021vit, liu2021swin}.
Following \cite{child2021vdvae}, we put the patch embedding layer after a residual block to form a downsampling operator.

For upsampling (bottom pane of Fig.~\ref{fig:qarv_method_architecture_down_up}), we use a 1x1 convolution layer followed by pixel shuffling, which is also known as sub-pixel convolution~\cite{shi2016subpixel}.
One can view this operation as an inverse of the patch embedding layer.
The sub-pixel convolution is sandwiched between two residual blocks for additional non-linearity (since the sub-pixel convolution alone is purely linear).


The residual block architecture is shown in the left pane of Fig.~\ref{fig:qarv_method_architecture_res_adaln}.
Following our previous work~\cite{duan2023qres}, we adopt ConvNeXt~\cite{liu2022convnext} as our residual block architecture in favor of its efficiency. 
In~\cite{duan2023qres}, the ConvNeXt block is used as is, and the entire model is used for fixed-rate compression.
In this paper, we modify the block architecture to make all residual blocks conditional on the input $\lambda$ for variable-rate compression.
We replace the Layer Normalization~\cite{ba2016layernorm} operation in the original ConvNeXt block with our proposed adaptive Layer Normalization (AdaLN) operation, which is described in detail later in Sec.~\ref{sec:qarv_method2_rate_conditional}.

Note that the choice of the residual block used in our model is not restricted to ConvNeXt.
Indeed, any block architecture that involves a Layer Normalization can be used with our method by replacing its Layer Normalization with AdaLN.
In Sec.~\ref{sec:qarv_exp_additional}, we present experiments where we use our methods with the Neighborhood Attention Transformer block~\cite{hassani2022neighborhood} as well.


\subsection{Rate-Conditional Mechanism}
\label{sec:qarv_method2_rate_conditional}
We achieve variable-rate compression by conditioning all residual blocks in the model on the R-D Lagrange multiplier $\lambda$. This is done by two modules: a $\lambda$ embedding network (on top of Fig.~\ref{fig:qarv_method_architecture_overall}) and the AdaLN (right pane of Fig.~\ref{fig:qarv_method_architecture_res_adaln}).
We explain them one by one in this subsection.

\textbf{The $\lambda$ embedding network} takes $\lambda$, a positive real number, as input and produces $e_\lambda$, a vector carrying information about $\lambda$.
Given an input $\lambda$, we first scale it using a natural logarithmic function to scale it into the log space.
The scaled value is then converted into a vector using a sinusoidal embedding operation, \ie, the method used for positional encoding in Transformers~\cite{vaswani2017attention} and time embedding in diffusion probabilistic models~\cite{ho2020ddpm}.
Finally, a multilayer perceptron (MLP) is applied to obtain $e_\lambda$.
The embedding $e_\lambda$ is used in all residual blocks to make the entire model conditional on $\lambda$.

\textbf{Adaptive Layer Normalization (AdaLN):}
the AdaLN module is built on the Layer Normalization~\cite{ba2016layernorm} operation.
In the context of image features, it normalizes (by subtracting the mean and dividing by the standard deviation) an image feature w.r.t. the channel dimension, typically followed by an affine transform with learned weights and biases.
In AdaLN (right pane of Fig.~\ref{fig:qarv_method_architecture_res_adaln}), we parameterize the weights and biases of this affine transform as the output of a simple network whose input is $e_\lambda$.
One can view this as using a hyper-network to produce the affine weights and biases of the Layer Normalization operation.
By doing this, the channel-wise mean and standard deviation of the output of AdaLN is determined by $\lambda$, making all residual blocks in the model (and thus all posteriors $\qzi$ and priors $\pzi$) conditional on $\lambda$.
We empirically found this AdaLN operation works well for variable-rate compression, being able to cover a wide range of rates (from around 0.2 to around 2.0 bits per pixel).


\textbf{Training strategy:}
Training is done by minimizing the variable-rate loss function (Eq.~\eqref{eq:qarv_method_loss_var_rate}).
Recall that one needs to specify $p_\Lambda$ used during training manually.
We heuristically define $p_\Lambda$ such that $\Lambda$ is the cube of a uniform random variable:
\begin{equation}
\label{eq:qarv_method2_lmb_training}
    \Lambda = \Delta^3 \text{ where } \ \Delta \sim U( \lambda_\text{low}^{1/3}, \lambda_\text{high}^{1/3} ),
\end{equation}
and $\lambda_\text{low}$ and $\lambda_\text{high}$ are the lower and upper boundary of the support of $\Lambda$.
In our experiments, we set $\lambda_\text{low} = 16$ and $\lambda_\text{high} = 2048$.
With Eq.~\eqref{eq:qarv_method2_lmb_training}, the pdf $p_\Lambda$ has a closed form expression:
\begin{equation}
p_\Lambda(\lambda) = \begin{cases}
    \frac{1}{\lambda_\text{high}^{1/3} - \lambda_\text{low}^{1/3}} \cdot \frac{1}{3} \lambda^{-2/3}  & \text{ if } \lambda_\text{low} \le \lambda \le \lambda_\text{high}
    \\
    0 & \text{ otherwise. }
\end{cases}
\end{equation}
Our choice of $p_\Lambda$ is motivated by an observation that uniform sampling in the cube root space of $\Lambda$ leads to a near-uniform sampling in the rate space (details in Appendix~\ref{sec:qarv_appendix_var_rate_training}).
Note that our QARV framework is not restricted to this specific choice of $p_\Lambda$.
Any probability density function, such as ones with different supports, can be used in place of $p_\Lambda$.

\textbf{Testing (compression and decompression):}
When compressing an image, we are able to adjust the R-D trade-off by varying the $\lambda$ input to the model. Note that the exact same $\lambda$ has to be shared between the sender side and the receiver side; otherwise, there is a mismatch between encoding and decoding.
In the implementation, we achieve this by including the $\lambda$ (stored as a float32 value) used for encoding in the bitstream.
As $\lambda$ consumes only a constant rate overhead (32 bits) regardless of the image resolution, the rate overhead caused by $\lambda$ becomes negligible when compressing high-resolution images.

\zhihao{
\textbf{Relationship to scalable coding:}
QARV employs multiple latent variables that retain an autoregressive relationship, so it supports scalable coding~\cite{radha2001mpeg4_scalable, ma2022deepfgs} by progressively sending the latent variables.
However, it is worth noting that this scalable coding strategy is independent of the proposed variable-rate compression method, in which all latent variables are used.
Since this paper focuses on (non-scalable) variable-rate compression, without otherwise specified, all following experiments are for non-scalable coding, except in Sec.~\ref{sec:qarv_exp_progressive} where we use scalable coding only to analyze the latent variables in our model.
}

\section{Experiments} \label{sec:qarv_experiments}

\begin{figure*}[ht]
    \centering
    \subfloat[]{
        \includegraphics[width=0.322\linewidth]{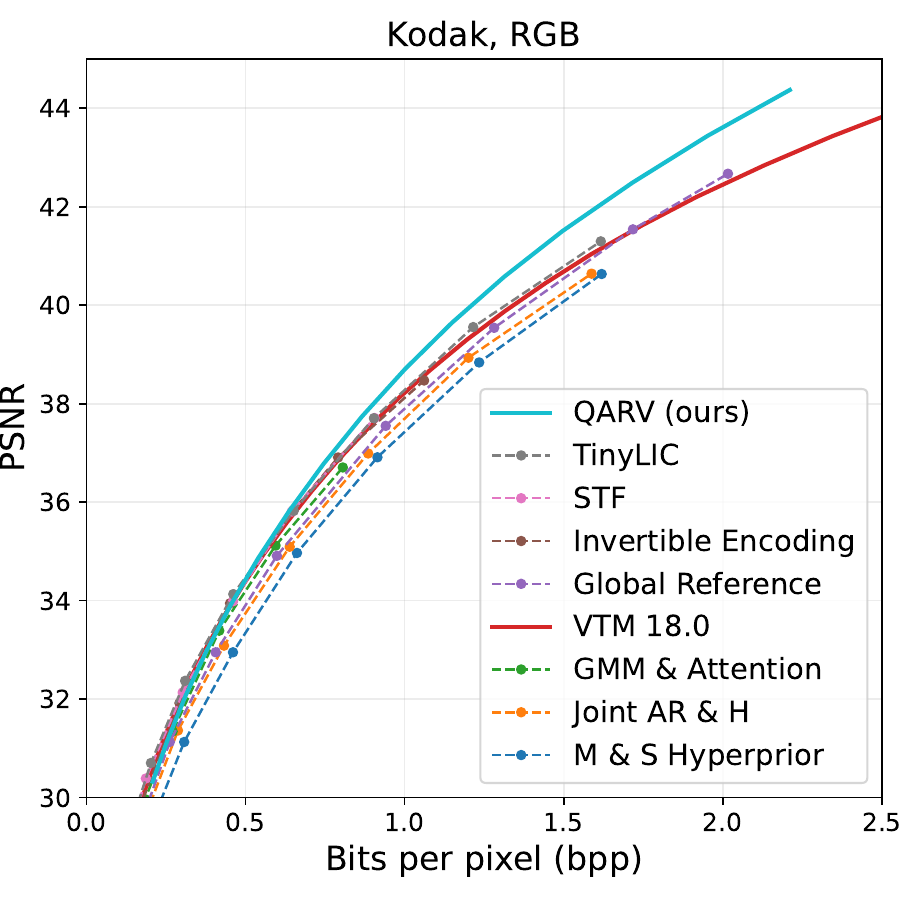}
        \label{fig:qarv_exp_main_kodak}
    }
    \subfloat[]{
        \includegraphics[width=0.322\linewidth]{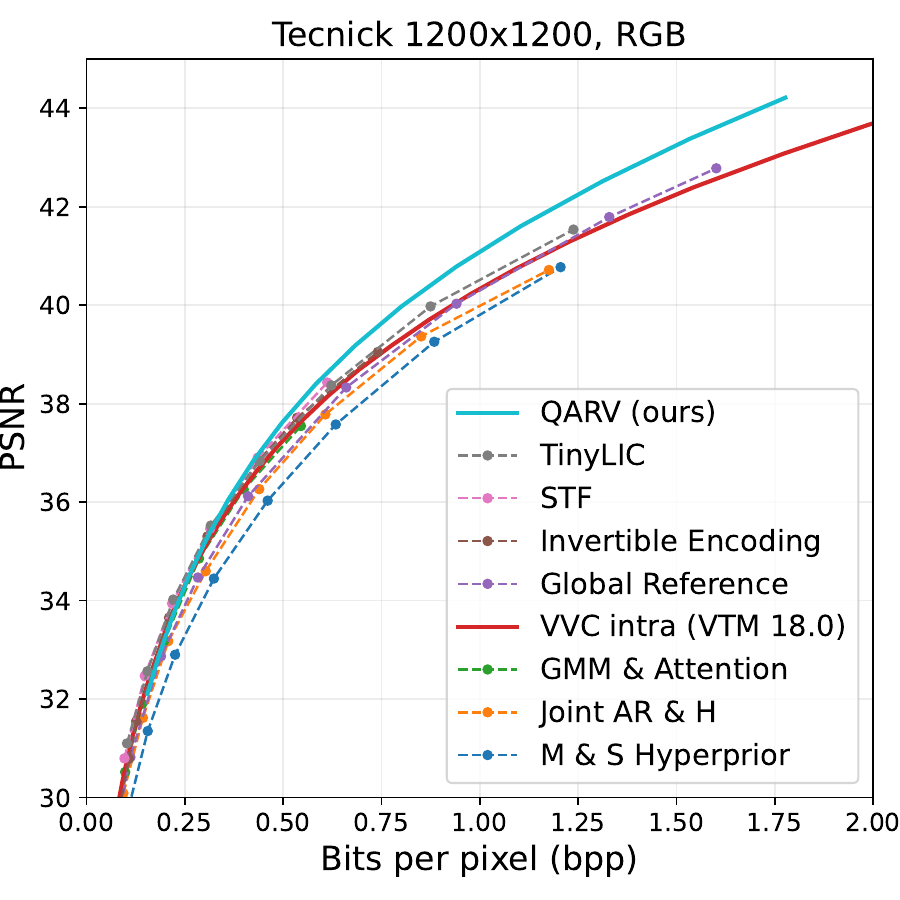}
        \label{fig:qarv_exp_main_tecnick}
    }
    \subfloat[]{
        \includegraphics[width=0.322\linewidth]{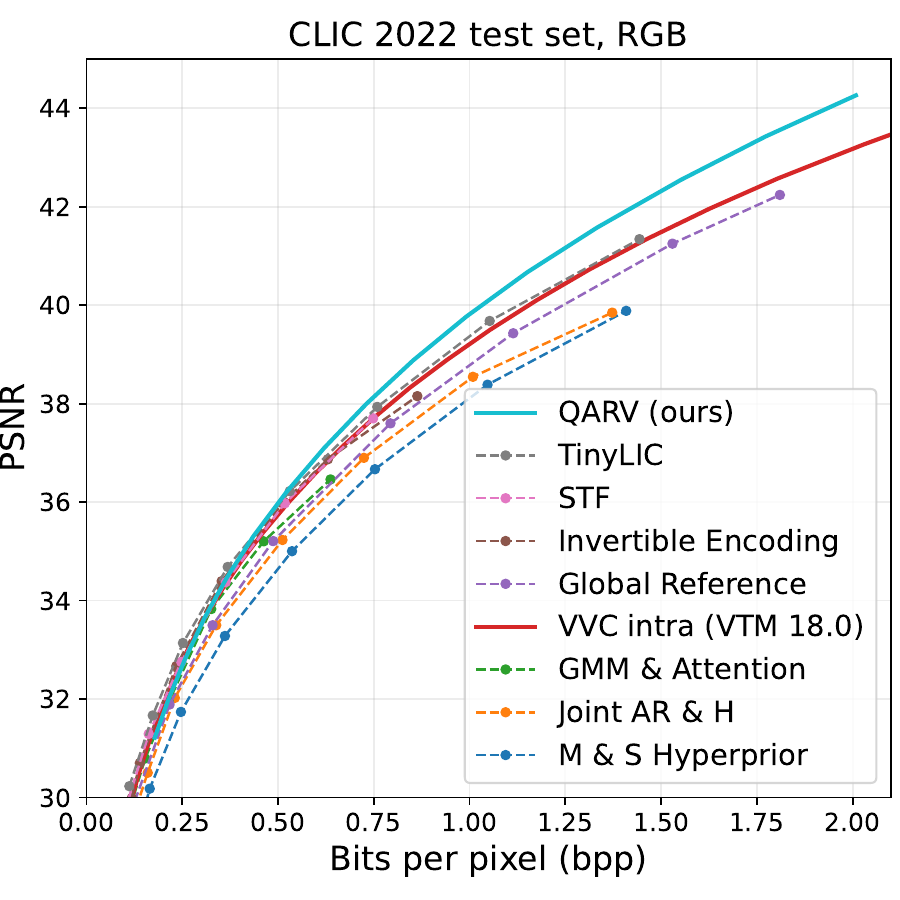}
        \label{fig:qarv_exp_main_clic}
    }
    \hfill
    \caption{\textbf{PSNR-bpp results for three test sets:} (a) Kodak, (b) Tecnick, and (c) CLIC 2022 test split. Dashed lines are fixed-rate methods (\ie, separate models for different rates), and solid lines are variable-rate methods (\ie, a single model or program with adjustable rates). Our QARV model, which supports variable-rate compression, achieves comparable PSNR with previous methods at low rates ($\leq$ 0.5 bpp) while outperforming them by a clear margin at high rates ($>$ 0.75 bpp).}
    \label{fig:qarv_exp_main}
\end{figure*}

\begin{table*}[ht]
\centering
\caption{Computational Complexity and BD-Rate of QARV Compared to Existing Learning-based Methods.}
\vspace{-0.2cm}
\begin{adjustbox}{width=\linewidth}
\begin{tabular}{l|cc|c:c|c:c|ccc:c}
\hline
                                             & &               & \multicolumn{2}{c|}{\textbf{Latency (CPU)}} & \multicolumn{2}{c|}{\textbf{Latency (GPU)}} & \multicolumn{4}{c}{\textbf{BD-rate (\%) w.r.t. VTM 18.0}} \\
                                             & \textbf{\# Models} & \textbf{\# Params.} & \multicolumn{1}{|c}{\textbf{Enc.}} & \multicolumn{1}{c|}{\textbf{Dec.}} & \multicolumn{1}{|c}{\textbf{Enc.}} & \multicolumn{1}{c|}{\textbf{Dec.}} & \textbf{Kodak}     & \textbf{Tecnick}   & \multicolumn{1}{c}{\textbf{CLIC}} & \multicolumn{1}{c}{\textbf{Avg.}}     \\ \hline
\textit{Variable-rate models}                &   &           &        &        &        &        &        &        &        &          \\
QARV (ours)                                  & 1 & \underline{93.4M}     & 0.757s & \underline{0.288s} & 0.211s & 0.096s & \underline{-5.899} & \underline{-8.938} & \underline{-6.951} & \underline{-7.263}   \\
\hline
\textit{Parallel context models}             &   &           &        &        &        &        &        &        &        &          \\
TinyLIC~\cite{lu2022tinylic}                 & 8 & 227M      & 2.410s & 2.489s & 0.237s & 0.316s & -3.100 & -5.059 & -5.573 & -4.578   \\
STF~\cite{zou2022stf}                        & 6 & 599M      & 0.644s & 0.740s & 0.139s & 0.177s & -2.092 & -6.219 & -2.479 & -3.597   \\
M \& S Hyperprior~\cite{minnen2018joint}     & 8 & 98.4M     & \underline{0.194s} & 0.360s & \underline{0.053s} & \underline{0.046} & 21.99  & 22.06  & 27.56  & 23.87    \\ \hline
\textit{Non-parallel AR models}              &   &           &        &        &        &        &        &        &        &          \\
Invertible Encoding~\cite{xie2021invertible} & 8 & 400M      & 2.324s & 5.263s & 2.170s & 4.844s & 0.338  & -1.510 & -2.154 & -1.109   \\
GMM \& Attention~\cite{cheng2020cvpr}        & 6 & 115M      & 1.859s & 4.921s & 2.276s & 4.978s & 4.552  & 1.378  & 3.489  & 3.140    \\
Global Reference~\cite{qian2021global_ref}   & 9 & 324M      & 251.8s & 259.6s & 41.52s & 49.76s & 6.578  & 4.648  & 8.485  & 6.570    \\
Joint AR \& H~\cite{minnen2018joint}         & 8 & 158M      & 4.133s & 7.322s & 2.147s & 5.169s & 11.51  & 9.375  & 14.09  & 11.66    \\ \hline
\end{tabular}
\end{adjustbox}
\begin{tablenotes}
    \footnotesize
    \item Testing environment: Intel 10700K CPU (using 8 threads), Nvidia 1080 Ti GPU, Linux system, PyTorch 1.13, CUDA 11.7.
    \item Latency is the average time to encode/decode a Kodak image (768$\times$512 pixels), averaged over all 24 images. Time includes entropy coding.
\end{tablenotes}
\label{table:qarv_exp_complexity_bdrate}
\end{table*}

We conduct extensive experiments to compare QARV against existing methods as well as analyze the components of QARV.
We begin by summarizing the experiment setting in Sec.~\ref{sec:qarv_exp_setting}, and we present the comparison with existing methods in Sec.~\ref{sec:qarv_exp_sota_comparison}, additional experiments in Sec.~\ref{sec:qarv_exp_additional}, and qualitative analysis in Sec.~\ref{sec:qarv_exp_progressive}.

\subsection{Experiment Settings}
\label{sec:qarv_exp_setting}
\textbf{Datasets:} we use the COCO~\cite{lin2014coco} dataset \textit{train2017} split to train our model. It contains 118,287 images with around $640\times420$ pixels. We randomly crop the images to $256 \times 256$ patches during training. We evaluate our model on three commonly-used test sets:
\begin{itemize}
    \item \textbf{Kodak}~\cite{kodak}. This dataset contains 24 images with either $512\times 768$ or $768\times 512$ pixels.
    \item \textbf{Tecnick TESTIMAGES}~\cite{asuni2014tecnick}. Following previous works~\cite{balle18hyperprior, minnen2018joint}, we use the \textit{RGB\_OR\_1200x1200} split, which contains 100 images with $1,200 \times 1,200$ pixels. We refer to this test set as \textbf{Tecnick} for short.
    \item \textbf{CLIC 2022 test set}. This is the latest test set from the Challenge on Learned Image Compression (CLIC)\footnote{http://compression.cc}. It contains 30 images with around $2048\times 1365$ pixels. In this paper, we refer to this test set as \textbf{CLIC}.
\end{itemize}

\zhihao{
\textbf{Metrics:} We use standard metrics to quantify rate and distortion.
The reconstruction distortion is measured by the peak signal-to-noise ratio (PSNR, higher is better):
\begin{equation}
\text{PSNR} \triangleq -10 \cdot \log_{10} \text{MSE},
\end{equation}
where pixel values are between 0 and 1, and the MSE is measured in the RGB space.
We measure data rate by bits per pixel (bpp, lower is better):
\begin{equation}
\text{bpp} \triangleq \frac{\text{\# of bits}}{\text{\# of image pixels}}.
\end{equation}
To obtain the overall PSNR and bpp for an entire dataset, we first compute PSNR and bpp for each image and then average them over all images.
We also use the BD-rate metric~\cite{bjontegaard2001bdrate} to measure the average rate difference w.r.t. an anchor method (lower is better).
We set the VVC intra codec~\cite{pfaff2021vvc_intra} as the BD-rate anchor in all our experiments.
}

\textbf{Training details:}
For the main experiment in Sec.~\ref{sec:qarv_exp_sota_comparison}, we train our QARV model on COCO images for 2 million iterations with a batch size of 32.
For all other experiments, including ablation studies, we train all models for 500k iterations with a batch size of 16.
Other hyperparameters (\eg, the optimizer and learning rate) stay the same across all experiments.
We list the full training details in Appendix~\ref{sec:qarv_appendix_hyp_param}.

\subsection{Comparison with Existing Methods}
\label{sec:qarv_exp_sota_comparison}
We first compare QARV with existing open-source methods for lossy image compression.
We use the recently developed VVC intra codec~\cite{pfaff2021vvc_intra} (VTM version 18.0) as the BD-rate anchor.
We list the details of all baseline methods, including the source of implementation and how we test them, in Appendix~\ref{sec:qarv_appendix_baseline_methods}.

The PSNR-bpp curves for the three test sets are shown in Fig.~\ref{fig:qarv_exp_main}.
Note in particular that most existing methods only support fixed-rate compression, while QARV is a variable-rate model.
We observe that at lower rates ($\leq 0.5$ bpp), most methods are comparable to each other, as the PSNR-bpp curves are highly overlapped.
However, starting at around $0.75$ bpp, our method outperforms other competitors by a clear margin, and the margin tends to grow larger as the rate increases.
This observation is consistent across all three test sets as well as in our previous work~\cite{duan2023qres}.
We attribute this bias toward higher rates to the fact that QARV, which has nine stochastic layers, contains far more latent variables than previous approaches, most of which have only two stochastic layers.
More latent variables enable better coding of pixel details, which largely benefits high-rate compression.

We report computational complexities and BD-rates of QARV against existing learning-based image codecs in Table~\ref{table:qarv_exp_complexity_bdrate}.
The total number of parameters is summed over all separately trained models, and the encoding/decoding latencies are measured for the highest rate (to represent the worst-case scenario).
In terms of the total number of parameters, QARV is more efficient than other methods due to the fact that it uses only a single model for all rates, while fixed-rate methods employ separate models for different rates.
In terms of CPU encoding latency, QARV is comparable with most existing methods within an order of magnitude.
In terms of CPU decoding latency, however, QARV visibly outperforms all compared methods.
Even when compared with the fastest baseline method, Mean \& Scale Hyperprior~\cite{minnen2018joint}, our approach saves 20\% of decoding time on CPU (0.288s vs. 0.360s).
This feature of fast decoding is favorable in many real-world applications where decoding speed is critical.
The results for GPU encoding/decoding latency are similar to the ones for CPU. 
Our model is efficient in decoding latency, being faster than most of the compared methods.
In the last columns of Table~\ref{table:qarv_exp_complexity_bdrate}, we report the BD-rate w.r.t. VTM version 18.0 (lower is better).
We observe that our approach achieves a better compression efficiency than all other approaches.
Overall, our QARV model is strong in both decoding latency and BD-rate, making it more appealing in practice than the compared baseline methods.

\begin{figure}[t]
    \centering
    \includegraphics[width=0.64\linewidth]{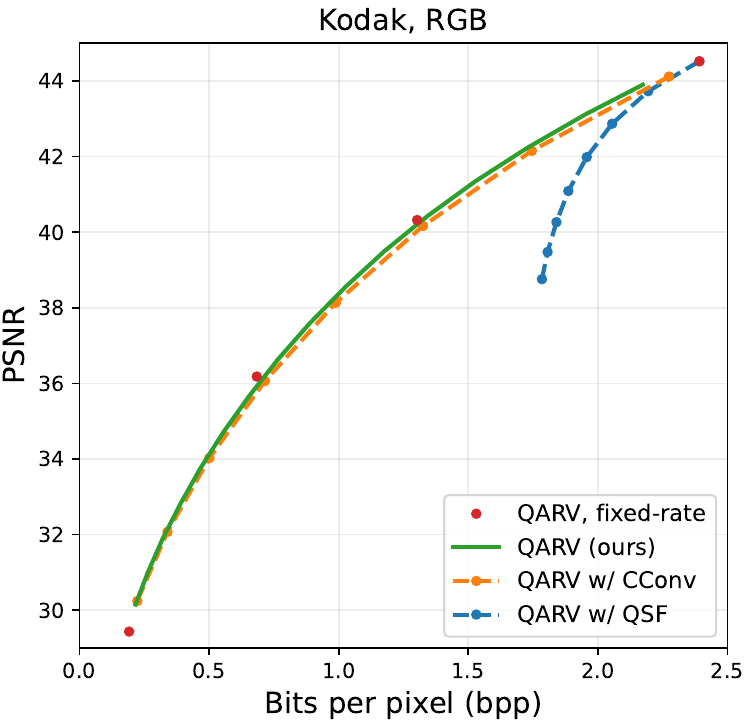}
    \vspace{-8pt}
    \caption{Comparison of different variable-rate compression methods for QARV. CConv is the conditional convolution operation~\cite{sohn2015conditional_vae}, and QSF is the quality scaling factor method~\cite{chen2020var_rate_quality_scaling}.
    }
    \label{fig:qarv_exp_var_rate}
\end{figure}

\subsection{Experimental Analysis: Variable-Rate Compression}
\label{sec:qarv_exp_additional}
In this section, we conduct experiments to analyze the variable-rate compression method of QARV. 
All BD-rates reported in this section are w.r.t. VTM version 18.0.

\textbf{Variable-rate compression methods:}
Fig.~\ref{fig:qarv_exp_var_rate} shows the comparison between our variable-rate method against previous methods when applied to QARV.
We first train a fixed-rate variant of QARV (\ie, without the $\lambda$ embedding module and AdaLNs) as the baseline, shown as \textit{QARV fixed-rate} in the figure.
The $\lambda$ values for the fixed-rate QARV are
\begin{equation}
    \lambda \in \{ 16, 128, 512, 2048 \}.
\end{equation}
Then, we apply two representative variable-rate methods: the quality-scaling factor~\cite{chen2020var_rate_quality_scaling} (QSF) and the conditional convolution~\cite{choi2019var_rate_conditional_ae} (CConv).
Both methods require a finite set of $\lambda$, which we choose to be
\begin{equation}
    \lambda \in \{ 16, 32, 64, 128, 256, 512, 1024, 2048 \}.
\end{equation}
QSF takes a pre-trained model (in our case, the one with $\lambda=2048$) and learns a scaling factor for each latent variable to adapt the model to an unseen $\lambda$.
As shown in the figure, \textit{QARV w/ QSF} produces much lower PSNR than other methods, suggesting that a simple scaling of latent variables is not sufficient for effective variable-rate compression using QARV.
On the other hand, CConv accompanies each convolutional layer by a fully connected layer, which is used to produce a scaling vector and a shifting vector for the convolution layer output given $\lambda$. This effectively conditions all convolutional layers, and thus all posteriors and priors, on the input $\lambda$.
We implement QARV with CConv by replacing every depth-wise convolutional layer in the residual blocks with a depth-wise CConv.
This ensures that \textit{QARV w/ CConv} and QARV have the same number of $\lambda$-adaptive layers and thus a fair comparison. 
We observe that \textit{QARV w/ CConv} achieves better performance than the one with QSF but is still below the fixed-rate QARV by a small margin.
Finally, we show the results of our variable-rate method, \ie, the $\lambda$ embedding module along with AdaLN layers, denoted as \textit{QARV (ours)} in Fig.~\ref{fig:qarv_exp_var_rate}.
Our method achieves a continuously variable-rate compression and at the same time outperforms other baselines at all rates.
Our PSNR-bpp curve is close to the fixed-rate performance using only a single model, demonstrating the effectiveness of our introduced methods.

\begin{figure}[t]
    \centering
    \includegraphics[width=0.64\linewidth]{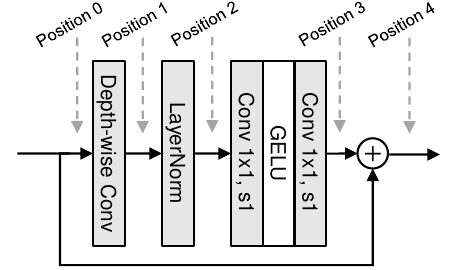}
    \caption{Illustration of possible positions in a residual block to insert the adaptive affine transform for variable-rate compression. Inserting the affine transform at position 2 is equivalent to using AdaLN.}
    \label{fig:qarv_exp_affine_positions}
\end{figure}

\textbf{Variants of AdaLN:}
Recall that AdaLN is a layer normalization operation followed by an adaptive affine transform.
We notice that the adaptive affine transform could instead be placed at any position in a residual block, as shown in Fig.~\ref{fig:qarv_exp_affine_positions}.
To verify our design of AdaLN, we train QARV with the affine transform placed at every possible position shown in the figure, and we report their performance in Table~\ref{table:qarv_exp_affine_positions}.
Note that placing the affine transform at \textit{position 1} is similar to using the CConv layer, except that here the input $\lambda$ can be continuously tuned thanks to the $\lambda$ embedding module.
Table~\ref{table:qarv_exp_affine_positions} shows that among all positions, position 2 (\ie, AdaLN) gives the best performance, reflecting the effectiveness of our design.
We heuristically conjecture that, since the output of Layer Normalization has a zero mean and unit standard deviation (channelwise), the affine transform at position 2 directly determines the mean and std, which presumably leads to more stable and efficient learning.

\begin{table}[t]
\centering
\caption{Comparison between Variants to the AdaLN.}
\vspace{-8pt}
\begin{adjustbox}{width=\linewidth}
\begin{tabular}{cc|c}
\hline
\textbf{Affine Transform Position} & \textbf{Normalization} & \textbf{Kodak BD-rate (\%)} \\ \hline
Position 0                         & Layer Norm             & 2.446            \\
Position 1                         & Layer Norm             & 2.005 \\
Position 2 (AdaLN, ours)           & Layer Norm             & \underline{-0.066}\\
Position 3                         & Layer Norm             & 1.970            \\
Position 4                         & Layer Norm             & 5.365            \\ \hline
Position 2                         & Group Norm             & 1.041            \\
Position 2                         & Instance Norm          & 10.26            \\ \hline
\end{tabular}
\end{adjustbox}
\label{table:qarv_exp_affine_positions}
\end{table}

\textbf{Adaptive normalization types:}
Instead of Layer Normalization, one could also use other types of normalization layers, such as Group Normalization~\cite{wu2018groupnorm} and Instance Normalization~\cite{huang2017adaptive_instance_norm}.
We experiment with Group/Instance Normalizations by using them in place of all Layer Normalizations in QARV while keeping all other settings unchanged.
Results are shown at the bottom part of Table~\ref{table:qarv_exp_affine_positions}.
Following modern image generative models~\cite{dhariwal2021diffusion_beat_gan}, we use a group size of 32 for Group Normalization.
In terms of BD-rate, we observe that Layer Normalization gives slightly better performance (by around 1.0\% BD-rate) than Group Normalization, both of which significantly outperform the Instance normalization (by around 10.0\% BD-rate).
We thus empirically conclude that Layer Normalization (and thus our AdaLN) is a relatively better choice for variable-rate compression than other types of normalization.





\begin{figure}[t]
    \centering
    \includegraphics[width=0.64\linewidth]{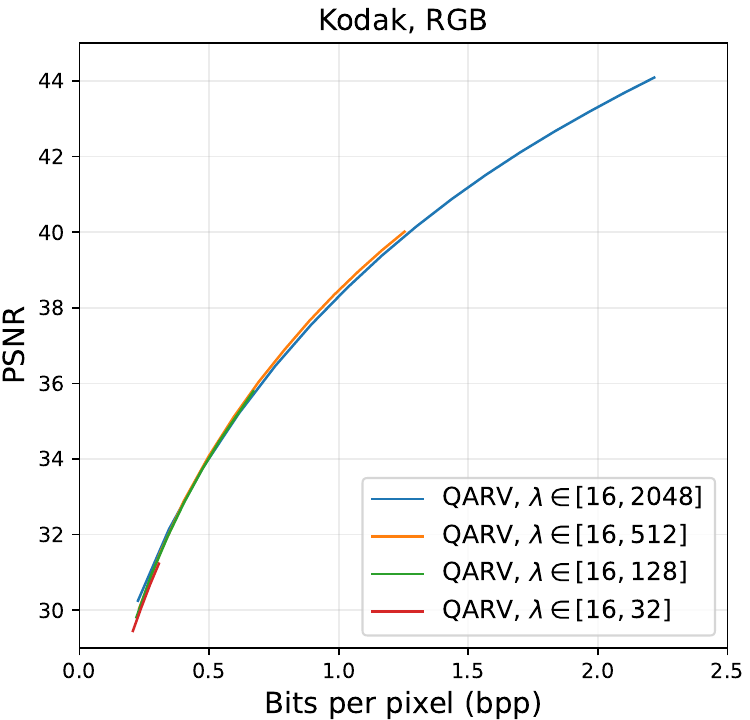}
    \vspace{-8pt}
    \caption{PSNR-bpp performance of QARV trained with different $\lambda$ ranges.
    We observe that QARV generalizes well to different rates.
    }
    \label{fig:qarv_exp_lmb_range}
\end{figure}

\textbf{Range of $\lambda$ during training:}
We train QARV with different $\lambda$ ranges and show R-D results in Fig.~\ref{fig:qarv_exp_lmb_range}.
We keep the lower bound $\lambda_\text{low}$ fixed to 16, and vary the upper bound $\lambda_\text{high}$ from $32$ to $2048$.
As $\lambda$ trades off rate and distortion, a larger range of $\lambda$ results in a wider R-D curve.
We first observe that, coarsely speaking, the R-D curves are mostly overlapped with each other.
Interestingly, when looking at the lowest rate (corresponding to $\lambda = 16$), the model with $\lambda \in [16,2048]$ gives the best R-D performance, indicating that training for high-rate compression also benefits low-rate compression.
Overall, we can conclude that QARV generalizes well to a wide range of rates.

\subsection{Experimental Analysis: Model Architecture}
\label{sec:qarv_exp_arch}
We conduct various experiments to analyze the architectural components of QARV.
All BD-rates reported in this section are w.r.t. VTM version 18.0.

\begin{figure}[t]
    \subfloat[Configuration A]{
        \includegraphics[width=0.31\linewidth]{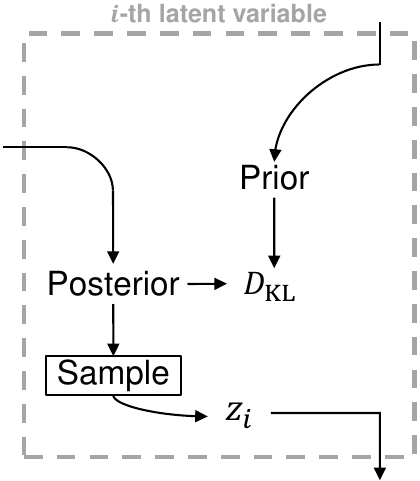}
        \label{fig:qarv_exp_hyp_vs_resvae_a}
    }
    \subfloat[Configuration B]{
        \includegraphics[width=0.31\linewidth]{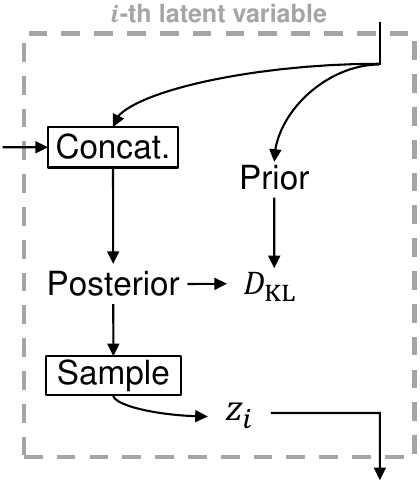}
        \label{fig:qarv_exp_hyp_vs_resvae_b}
    }
    \subfloat[Configuration C]{
        \includegraphics[width=0.31\linewidth]{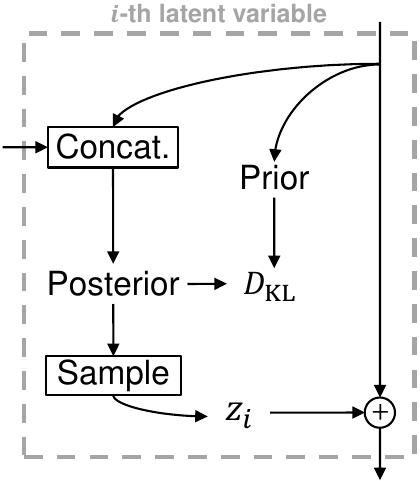}
        \label{fig:qarv_exp_hyp_vs_resvae_c}
    }
    \caption{
        Latent variable block configurations. Config A resembles the Hyperprior model~\cite{balle18hyperprior}. Config B incorporates bi-directional inference to A, and config C (which is used in QARV) incorporates a residual connection to B. Performance comparison is provided in Table~\ref{table:qarv_exp_hyp_vs_resvae}.
    }
    \label{fig:qarv_exp_hyp_vs_resvae}
\end{figure}
\begin{table}[t]
\centering
\caption{Comparison between Latent Variable Block Configurations.}
\vspace{-8pt}
\begin{tabular}{cc|c|c}
\hline
\textbf{Configuration} & $N$ & \textbf{Params.} & \textbf{Kodak BD-rate (\%)} \\ \hline
A                     & 2   & 29.8M   & 24.96              \\
B                     & 2   & 31.2M   & 18.20              \\
C                     & 2   & 31.2M   & \underline{14.06}  \\ \hline
A                     & 3   & 32.3M   & 22.51              \\
B                     & 3   & 34.1M   & 9.076              \\
C                     & 3   & 34.1M   & \underline{3.761}  \\ \hline
\end{tabular}
\label{table:qarv_exp_hyp_vs_resvae}
\end{table}

\textbf{Impact of bi-directional inference and residual coding:}
In this section, we conduct an ablation study to examine the two key features of QARV: bi-directional inference and residual coding, both of which are absent in the Hyperpiror model~\cite{balle18hyperprior}.
We first construct a small model with a latent variable block architecture that removes these two structures (config A, Fig.~\ref{fig:qarv_exp_hyp_vs_resvae_a}).
We also reduce the number of latent variables to $N=2$, where the two variables have the resolution of $64\times$ and $16\times$ spatially downsampled w.r.t. the original image.
This architecture resembles the two-layer Hyperprior model extensively used in existing learned image compression methods.
We train such a model with the same setting as before and show the results in the first row of Table~\ref{table:qarv_exp_hyp_vs_resvae}.
We then progressively incorporate bi-directional inference and residual connection into the latent variable block, which we refer to as config B (Fig.~\ref{fig:qarv_exp_hyp_vs_resvae_b}) and config C (Fig.~\ref{fig:qarv_exp_hyp_vs_resvae_c}), respectively.
Results for config B and C are included in Table~\ref{table:qarv_exp_hyp_vs_resvae}, where we can observe that each of them improves BD-rate by more than 4\%, and they maintain a comparable parameter size compared to config A.
The same set of experiments is conducted for $N=3$, where the latent variables are $64\times, 16\times,$ and $8\times$ spatially downsampled w.r.t. the original image.
Results are shown in the bottom segment of Table~\ref{table:qarv_exp_hyp_vs_resvae}.
The conclusion is consistent with when $N=2$, \ie, config C is better in terms of BD-rate than the other two configurations.
An interesting observation is that when compared to $N=2$, config A with $N=3$ only achieves around 2\% BD-rate improvement (22.51\% vs. 24.96\%), while both config B and config C achieve more than 9\% BD-rate improvements.
This suggests that bidirectional inference and residual connection are essential for improving the R-D performance when scaling up to a deep hierarchy of latent variables.

\textbf{Choices of the Residual Block:}
We experiment with other types of residual blocks in addition to the ConvNeXt block.
We adopt a recent Vision Transformer architecture, the Neighbourhood Attention Transformer (NAT)~\cite{hassani2022neighborhood}, to use in QARV.
Since a NAT block also contains Layer Normalizations, we replace the Layer Normalization in front of the MLP in each NAT block with AdaLN in the same way as we did for ConvNeXt blocks.
We show experimental results in Table~\ref{table:qarv_exp_res_block}, where for a fair comparison, we tune the number of channels of NAT blocks such that the total number of parameters approximately match the model with ConvNeXt blocks.
We observe that, with a comparable amount of parameters, the ConvNeXt block performs better than the NAT block by a margin of around 2\% BD-rate.
We thus use ConvNeXt as the choice of residual block in QARV.

\begin{table}[t]
\centering
\caption{Comparison of the Choices of Residual blocks.}
\vspace{-8pt}
\begin{tabular}{c|c|c}
\hline
\textbf{Residual Block Type} & \# \textbf{Params} & \textbf{Kodak BD-rate (\%)} \\ \hline
ConvNeXt~\cite{liu2022convnext} & 93.4M           & \underline{-0.066}           \\
NAT~\cite{hassani2022neighborhood} & 103.6M       & 2.916                       \\ \hline
\end{tabular}
\label{table:qarv_exp_res_block}
\end{table}

\begin{figure*}[ht]
    \centering
    \subfloat[\textbf{Progressive decoding:} The $i$-th image is decoded using $Z_{1:i}$, \ie, from the first latent variable up to $Z_i$.]{
        \includegraphics[width=0.98\linewidth]{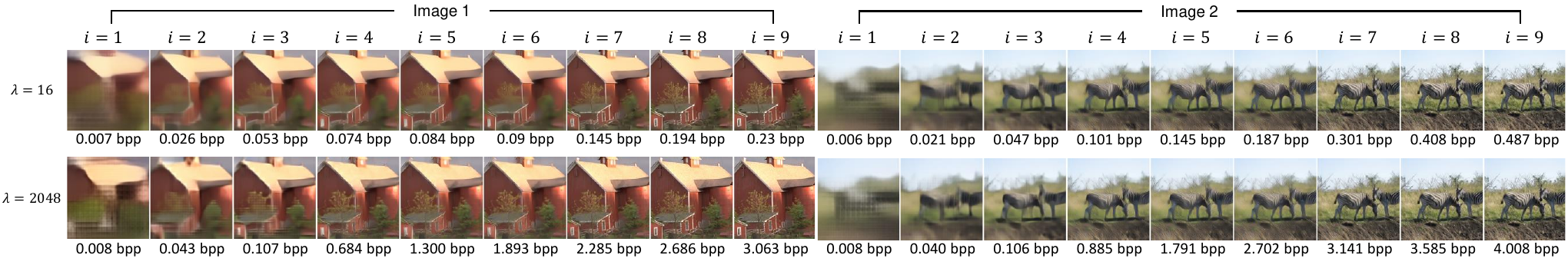}
        \label{fig:qarv_exp_decoding_progressive}
    }
    \vspace{4pt}
    \subfloat[
    \textbf{Leave-one-out decoding:} The $i$-th image is decoded using $Z_{1:N} \setminus {Z_i}$, \ie, all latent variables except $Z_i$.
    ]{
        \includegraphics[width=0.98\linewidth]{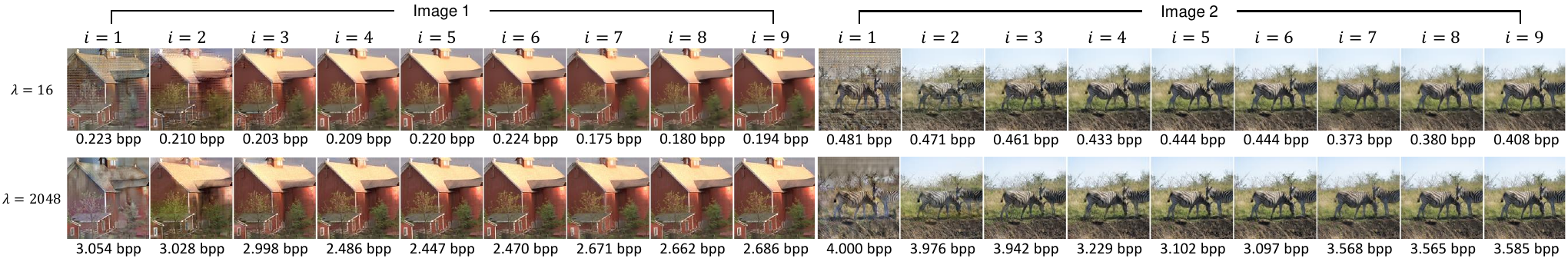}
        \label{fig:qarv_exp_decoding_exclude}
    }
    \vspace{4pt}
    \subfloat[
    \textbf{Independent decoding:} The $i$-th image is decoded using $Z_i$ only.
    ]{
        \includegraphics[width=0.98\linewidth]{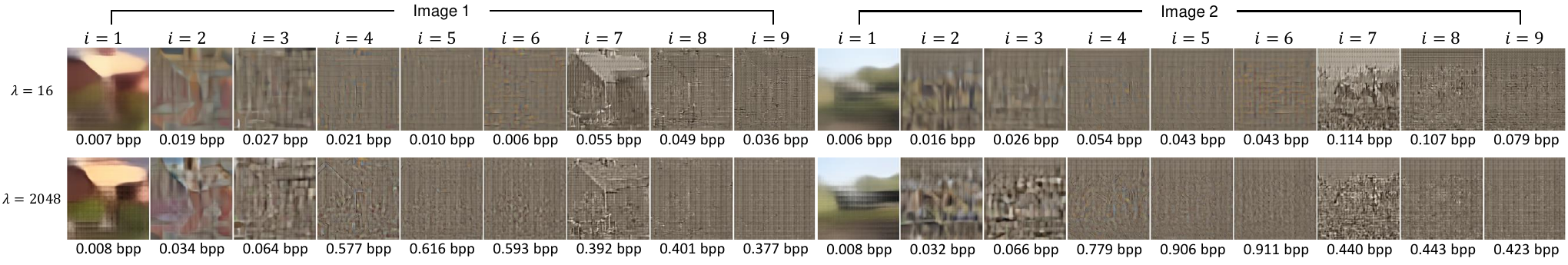}
        \label{fig:qarv_exp_decoding_single}
    }
    \hfill
    \vspace{-2pt}
    \caption{
    \textbf{Results of decoding only a subset of latent variables of an image.} We choose the subset in three different ways, shown in (a), (b), and (c), respectively.
    For each set of experiments, we show results for two images and two $\lambda$ values.
    The resulting images give hints on what information is carried by each latent variable.
    Image resolution is $256 \times 256$. Best if viewed by zooming in.
    }
    \label{fig:qarv_exp_progressive}
\end{figure*}

\subsection{Qualitative Analysis}
\label{sec:qarv_exp_progressive}
A key feature of the QARV model is that it compresses (and decompresses) images starting from low-dimensional latent variables to high-dimensional ones.
This leads to an interesting question: what part of the information of the uncompressed image does each latent variable carry?
In the following, we present several experiments to study this question, which leads to an unsurprising conclusion that QARV compresses images from global features to local features in a coarse-to-fine fashion.

\textbf{Progressive decoding:}
We start with progressively decoding the latent variables of an image, a common way to analyze the latent space of hierarchical VAEs.
Specifically, the image $X$ is first encoded into latent variables $Z_{1:N}$, and we decode the image using only the first subset of $i$ latent variables (for the remaining ones, we use their prior mean $\hat{\mu}_i$ instead).
We do this for $i \in \{ 1, 2, ..., N \}$ and show results in Fig.~\ref{fig:qarv_exp_decoding_progressive}, where several interesting observations can be made.
First, it is evident that the first latent variable, $Z_1$, encodes ``coarse features" of an image, such as brightness and color.
Note that $Z_1$ consumes only a very small number of bits compared to the remaining latent variables, as it has the lowest dimension (64x downsampled w.r.t. the original image).
Second, the high-dimensional latent variables carry mostly imperceptible details in the image.
For example, in the group of $\lambda=2048$ for image 1, the image with $i=6$ is almost visually indistinguishable from the image with $i=9$.
Lastly, the qualitative results suggest that the QARV model may be able to learn a semantic latent space.
As can be seen in the results for image 2, with around $0.1$ bpp, one could already recognize semantic information, such as the location of the sky, ground, and zebras.

\textbf{Leave-one-out decoding:}
In addition to progressive decoding, one can actually choose to decode an arbitrary subset of $Z_{1:N}$ to obtain a reconstruction.
We conduct experiments where we decode $Z_{1:N}$ except for a single $Z_i$, which we refer to as ``leave-one-out decoding", and show the results in Fig.~\ref{fig:qarv_exp_decoding_exclude}.
We observe that, for both images, leaving $Z_1$ or $Z_2$ out causes an obvious degradation in reconstruction quality, but leaving out other latent variables does not visibly impact the perceptual quality.
This suggests that the first subset of latent variables, despite that they only consume a small amount of rate, is most critical in reconstructing images.
From another perspective, QARV is potentially robust to packet loss in practical applications, given that the first few bitstreams are intact.

\textbf{Independent decoding}:
To further analyze the information contained in each latent variable, we decode each single $Z_i$ separately, the results of which are shown in Fig.~\ref{fig:qarv_exp_decoding_single}.
Observing the images corresponding to $i=1$, we again confirm that $Z_1$ contains the color and brightness information of images.
Interestingly, our model learns to represent color information only in the low-dimensional latent variables ($Z_1$ and $Z_2$) at a low rate, while reserving high-dimensional latent variables ($Z_{\ge 3}$) for almost purely luminance information, such as edges and other high-frequency details.
This coincides with the chroma subsampling strategy extensively used in traditional codecs, which is proven to be effective in improving R-D efficiency.
Devoting lower rates to color information also aligns with human perception, which is known to be much more sensitive to variations in brightness than color~\cite{van2001vision}.
Note that our model learns this rate allocation strategy without explicit supervision, but by merely learning to optimize rate and distortion.

To summarize, we find that QARV compresses and decompresses images in a coarse-to-fine manner, from high-level features to low-level ones.
We hypothesize that this coding scheme captures the hierarchical nature of images and human perception, and it could be used for multi-task processing in the compressed domain (\eg, using low-dimensional latent variables for classification while high-dimensional ones for segmentation).
However, applying recent approaches for compressed domain object recognition~\cite{xu2020learningfrequency, wang2022wacv_compressed, duan2023tcsvt} to a hierarchical architecture is non-trivial, and we leave it to future work.

\section{Conclusion}
We have shown that deep ResNet VAEs can be used for practical lossy compression by applying test-time quantization and quantization-aware training.
We present a new lossy image compression framework, QARV, along with a neural network implementation for it.
QARV provides a better compression efficiency than previous methods on three test sets, at the same time supporting variable-rate compression.
The progressive coding scheme of QARV better matches human perception and may lead to potential applications of compressed domain object recognition.
We believe the introduction of QARV provides valuable insights into future directions of lossy compression, as well as building deeper connections between data compression and generative modeling.


%



\ifCLASSOPTIONcaptionsoff
  \newpage
\fi



%



\newpage

\bibliographystyle{IEEEtran}
\bibliography{references}

%

\begin{IEEEbiography}[{\includegraphics[width=1in,height=1.25in,clip,keepaspectratio]{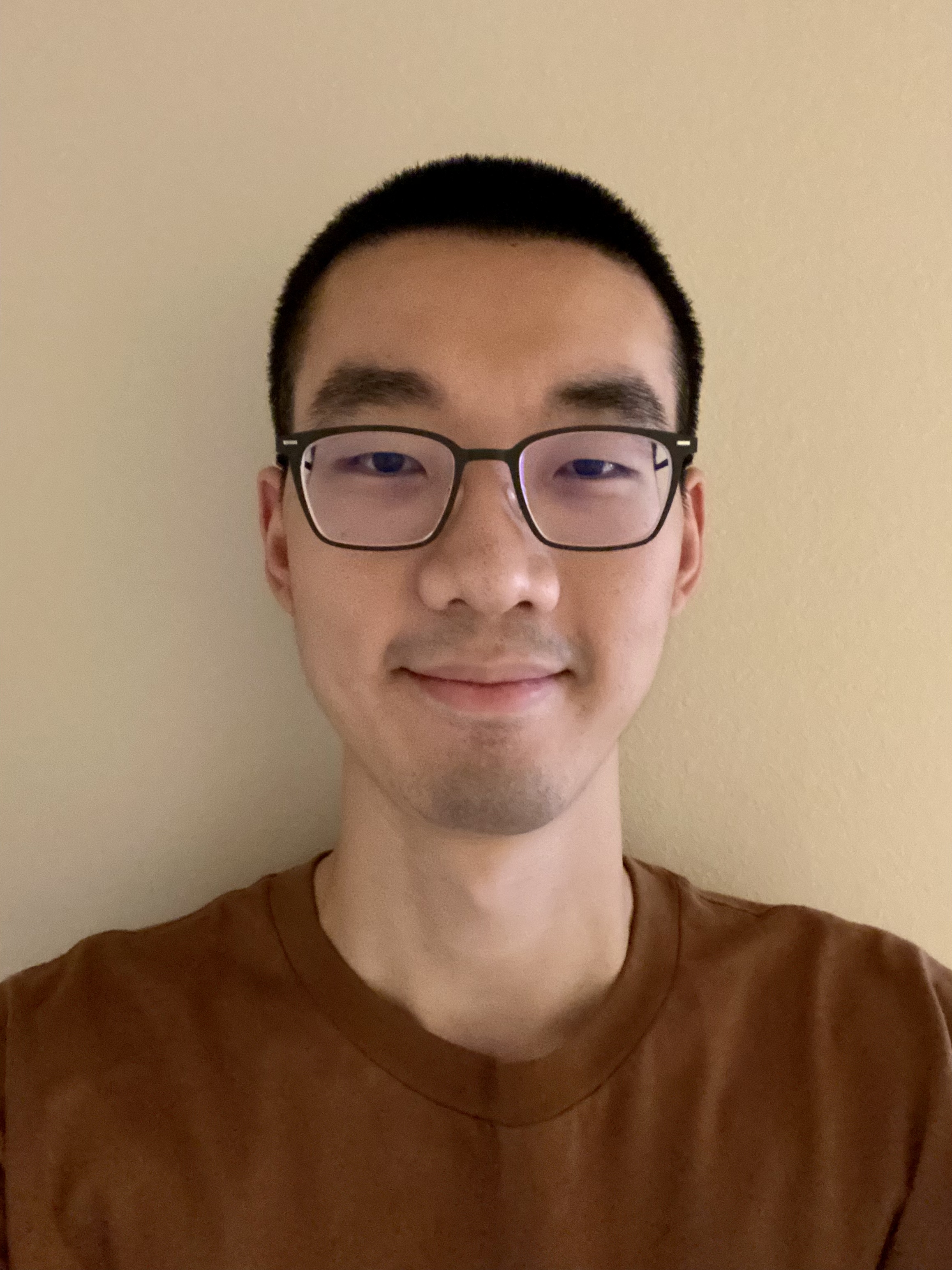}}]{Zhihao Duan}
received a B.E. degree in electrical engineering from Shanghai Jiao Tong University, Shanghai, China, in 2018 and an M.S. degree in electrical and computer engineering from Boston University, Boston, MA, USA, in 2020. He is now a Ph.D. candidate at the Video and Image Processing Laboratory at Purdue University, West Lafayette, IN, USA.
His research interests lie in the intersection of data compression and machine learning.
Recently, he received the 2022 IEEE PCS Best Paper Finalist and the 2023 IEEE WACV Best Algorithms Paper Award.
\end{IEEEbiography}

\begin{IEEEbiography}[{\includegraphics[width=1in,height=1.25in,clip,keepaspectratio]{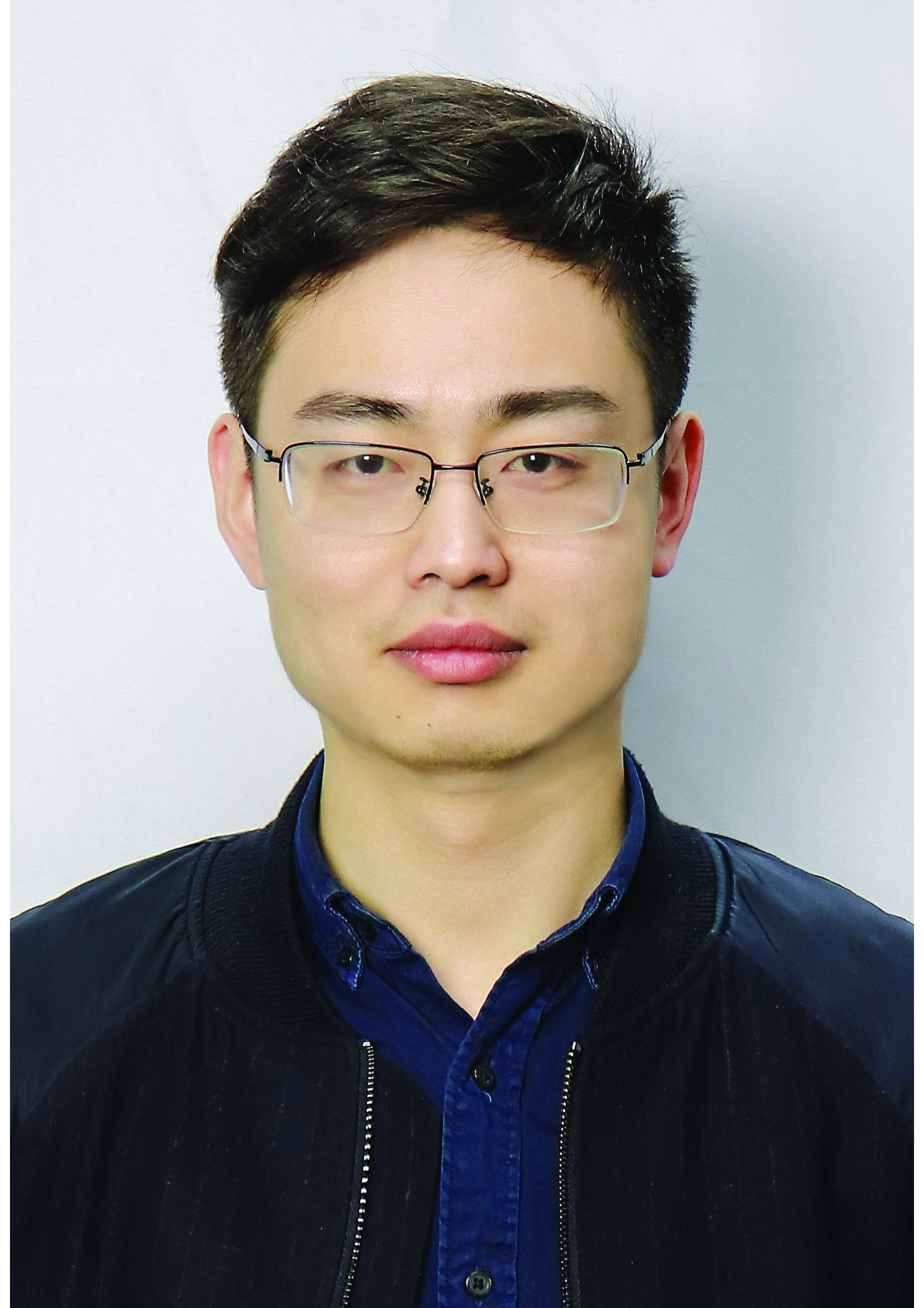}}]{Ming Lu} is an Associate Researcher in the School of Electronic Science and Engineering, Nanjing University, Jiangsu, China. He received his B.S. and Ph.D. degrees in the School of Electronic Science and Engineering from Nanjing University, Jiangsu, China, in 2016 and 2023 respectively. His current research focuses on deep learning-based image/video coding. He is a co-recipient of the 2018 ACM SIGCOMM Student Research Competition Finalist, the 2020 IEEE MMSP Image Compression Grand Challenge Best Performing Solution, and the 2023 IEEE WACV Best Algorithms Paper Award.
\end{IEEEbiography}

\begin{IEEEbiography}[{\includegraphics[width=1in,height=1.25in,clip,keepaspectratio]{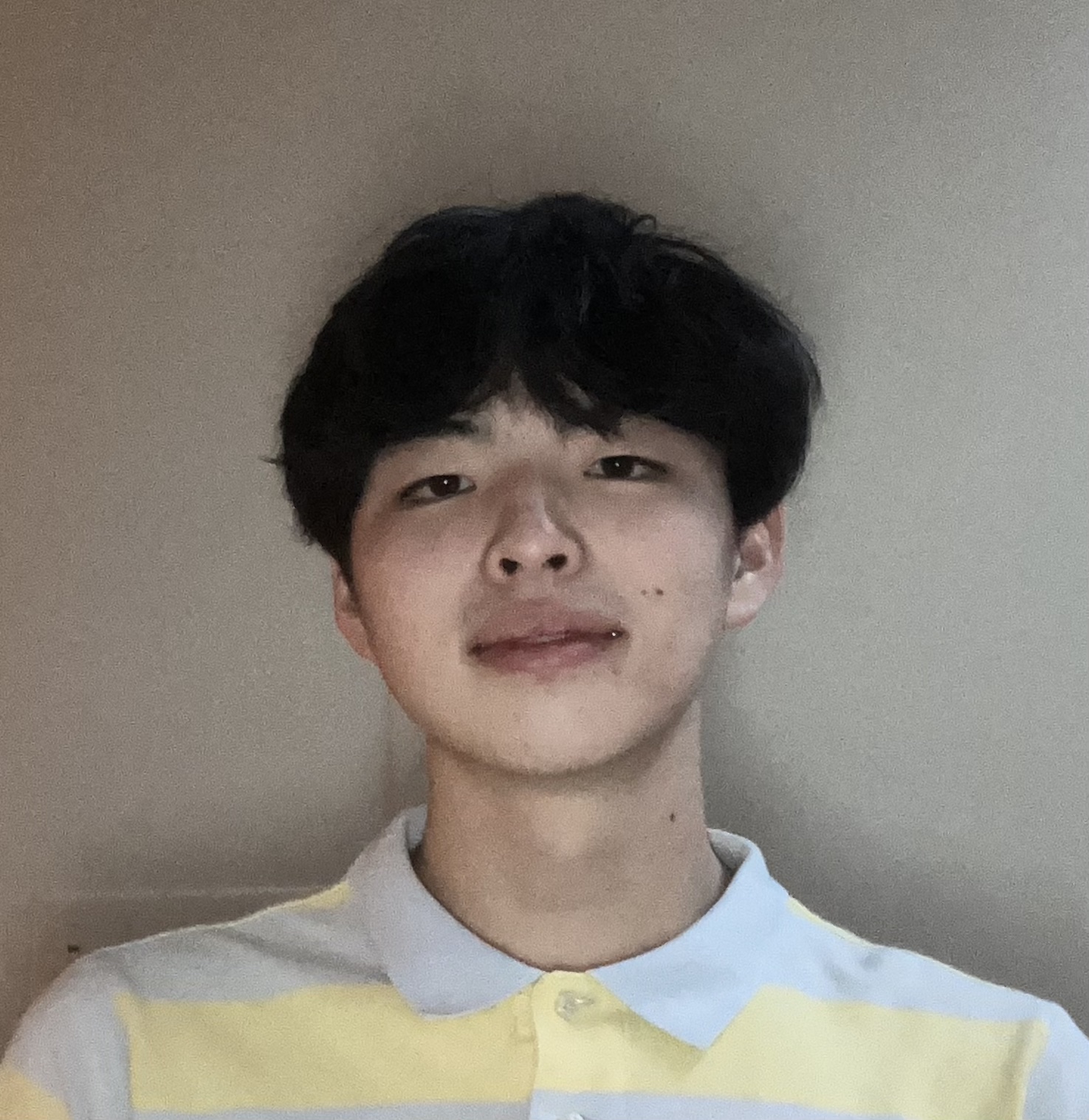}}]{Jack Ma} 
	is a non-degree student studying mathematics and computer science at Purdue University, West Lafayette, IN, USA. He currently conducts research at the Video and Image Processing Laboratory at Purdue University. His research interests include computer vision and image processing.
\end{IEEEbiography}

\begin{IEEEbiography}[{\includegraphics[width=1in,height=1.25in,clip,keepaspectratio]{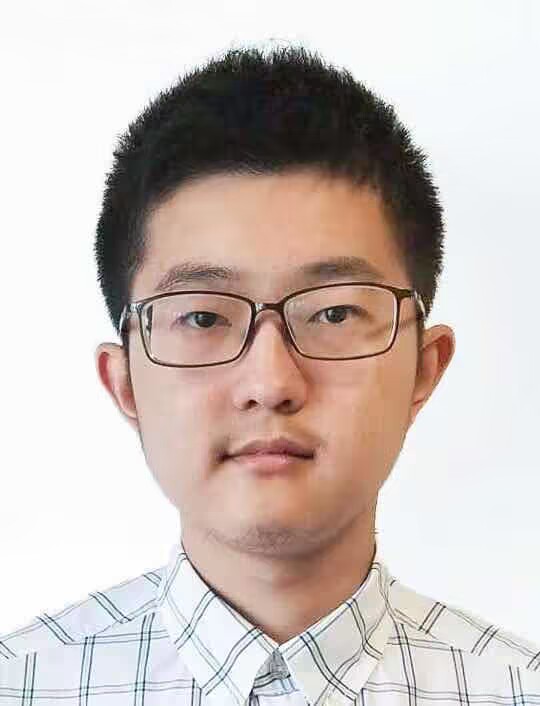}}]{Yuning Huang} 
	received a B.E. degree in Electronic and Computer Engineering from Zhejiang University, Hangzhou, China, in 2021. He is now pursuing a Ph.D. degree at the Video and Image Processing Laboratory at Purdue University, WestLafayette, IN, USA. His research interests include deep learning-based image/video enhancement and compression. 
\end{IEEEbiography}

\begin{IEEEbiography}[{\includegraphics[width=1in,height=1.25in,clip,keepaspectratio]{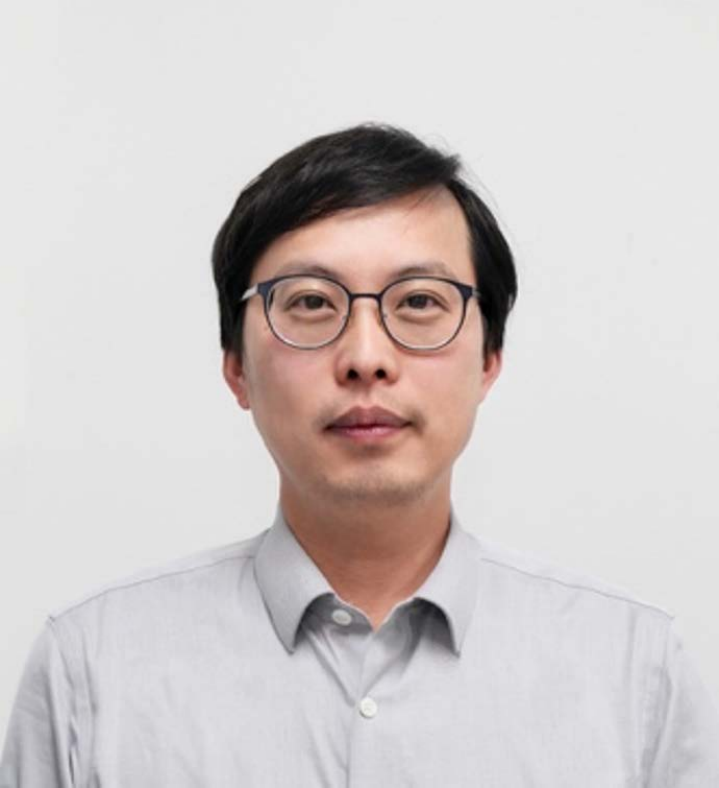}}]{Zhan~Ma} 
	is a Professor in the School of Electronic Science and Engineering, Nanjing University, Jiangsu, 210093, China. He received his Ph.D. from New York University, New York, in 2011 and his B.S. and M.S. from the Huazhong University of Science and Technology, Wuhan, China, in 2004 and 2006 respectively. From 2011 to 2014, he has been with Samsung Research America, Dallas, TX, and  Futurewei Technologies, Inc., Santa Clara, CA, respectively. His research focuses include learned image/video coding and computational imaging. He was awarded the 2018 PCM Best Paper Finalist, the 2019 IEEE Broadcast Technology Society Best Paper Award,  the 2020 IEEE MMSP Grand Challenge Best Image Coding Solution, the 2023 IEEE WACV Best Algorithms Paper Award, and the 2023 IEEE Circuits and Systems Society Outstanding Young Author Award.
\end{IEEEbiography}

\begin{IEEEbiography}[{\includegraphics[width=1in,height=1.25in,clip,keepaspectratio]{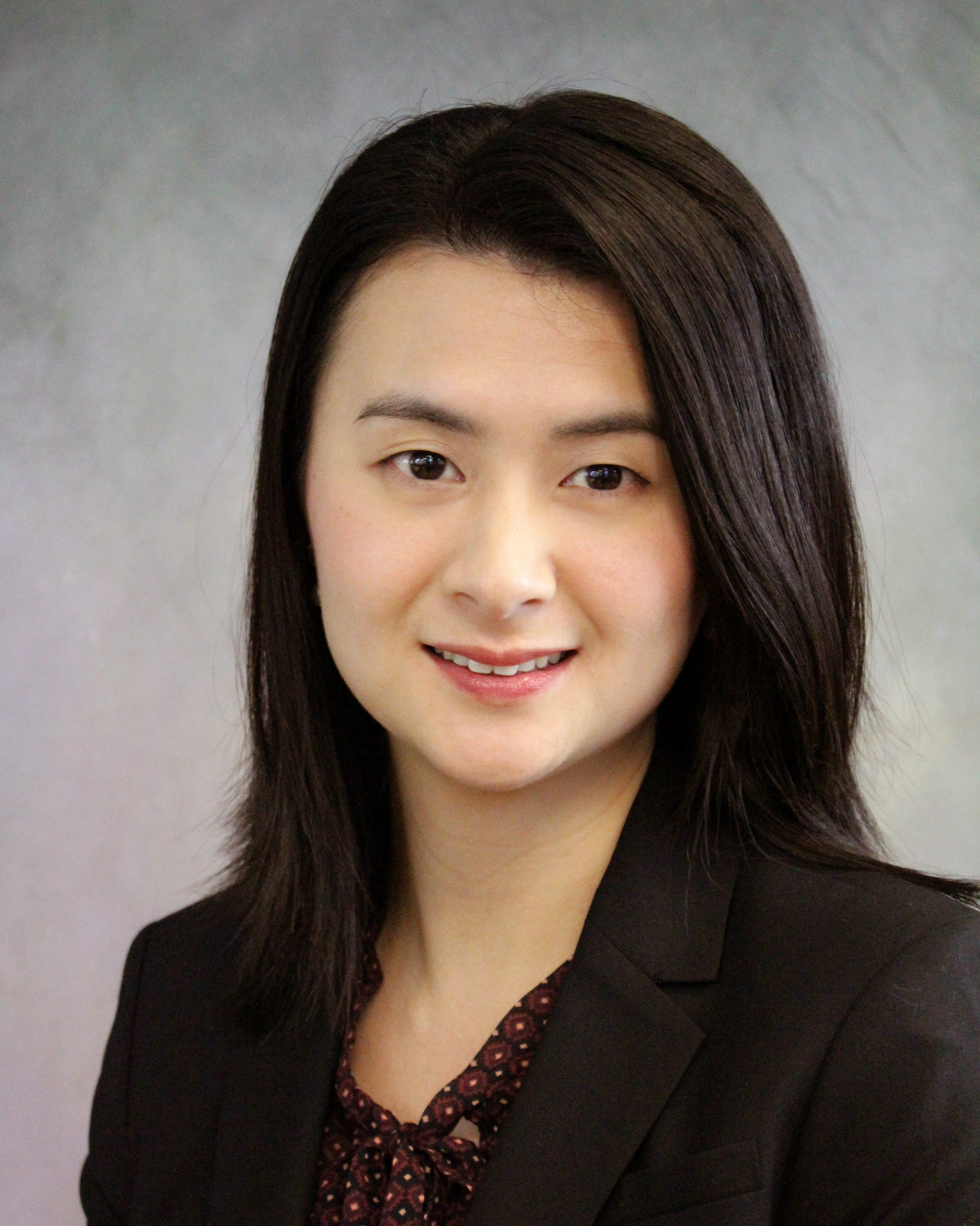}}]{Fengqing Zhu}
is an Associate Professor of Electrical and Computer Engineering at Purdue University, West Lafayette, Indiana. She received the B.S.E.E. (with highest distinction), M.S. and Ph.D. degrees in Electrical and Computer Engineering from Purdue University. From 2012 to 2015, she was a Staff Researcher with Futurewei Technologies, Inc., Santa Clara, CA, USA, where she received the Certification of Recognition for Core Technology Contribution in 2012. Her research interests include smart health with a focus on image based dietary assessment and wearable sensor data analysis, visual coding for machines, and application-driven visual data analytics. She is the recipient of an NSF CISE Research Initiation Initiative (CRII) award in 2017, a Google Faculty Research Award in 2019, and an ESI and trainee poster award for the NIH Precision Nutrition workshop in 2021.
\end{IEEEbiography}

\newpage
\appendices
\newpage



\section{Training Details}
\label{sec:qarv_appendix_hyp_param}
We list detailed training information in Table~\ref{table:qarv_appendix_hyp_param}, including data augmentation, hyperparameters, and training devices.
We use different settings for the main experiment and the additional experiments.
In the main experiment, we train our model until convergence, which requires around 10 days of training on a dual-GPU machine.
For ablation study experiments, we train our model with a smaller batch size (16 instead of 32) and a shorter training period (500k iterations instead of 2M iterations) to reduce training costs.

The training and testing results of QARV are fully reproducible using our open-source code.

\begin{table}[ht]
\small
\centering
\caption{Training Hyperparameters.}
\vspace{-0.2cm}
\begin{tabular}{l|c|c}
               & Main model         & Ablation study      \\ \hline
Training set   & COCO 2017 train    & COCO 2017 train     \\
\# images      & 118,287            & 118,287             \\
Image size     & Around 640x420     & Around 640x420      \\ \hdashline
Data augment.  & Crop, h-flip       & Crop, h-flip        \\
Train input size & 256x256          & 256x256             \\ \hdashline
Optimizer      & Adam               & Adam                \\
Learning rate  & $2 \times 10^{-4}$ & $2 \times 10^{-4}$  \\
LR schedule    & Constant + cosine  & Constant            \\
Weight decay   & 0.0                & 0.0                 \\ \hdashline
Batch size     & 32                 & 16                  \\
\# iterations  & 2M                 & 500K                \\
\# images seen & 64M                & 8M                  \\ \hdashline
Gradient clip  & 2.0                & 2.0                 \\
EMA            & 0.9999             & 0.9999              \\ \hdashline
GPUs           & 2 $\times$ RTX 3090 & 1 $\times$ Quadro 6000 \\
Time           & 260h               & 87h                 \\
\end{tabular}
\label{table:qarv_appendix_hyp_param}
\end{table}

\section{Baseline Methods}
\label{sec:qarv_appendix_baseline_methods}
\textbf{Learning-based image codecs:}
Recall that we adopt a number of learning-based image codecs to benchmark our method in Sec.~\ref{sec:qarv_exp_sota_comparison}.
In this section, we list the software implementations that we used to test these codecs.
For a fair comparison, all methods are implemented using the PyTorch~\cite{pytorch2019neural} library.

\begin{table}[ht]
\centering
\footnotesize
\caption{Baseline Methods.}
\begin{adjustbox}{width=\linewidth}
\begin{tabular}{c|l}
\hline
\textbf{Method}                             & \multicolumn{1}{c}{\textbf{Implementation}} \\ \hline
TinyLIC~\cite{lu2022tinylic}                & github.com/lumingzzz/TinyLIC                \\ \hline
STF~\cite{zou2022stf}                       & github.com/Googolxx/STF/issues              \\ \hline
M \& S Hyperprior~\cite{minnen2018joint}    & github.com/InterDigitalInc/CompressAI       \\ \hline
Invertible Encoding~\cite{xie2021invertible} & github.com/xyq7/InvCompress                 \\ \hline
GMM \& Attention~\cite{cheng2020cvpr}       & github.com/InterDigitalInc/CompressAI       \\ \hline
Global Reference~\cite{qian2021global_ref}  & github.com/damo-cv/img-comp-reference       \\ \hline
Joint AR \& H~\cite{minnen2018joint}        & github.com/InterDigitalInc/CompressAI       \\ \hline
\end{tabular}
\end{adjustbox}
\label{table:qarv_appendix_baseline_source}
\end{table}

\textbf{Versatile Video Coding (VVC):}
we also show the commands we used to test the VVC intra-prediction mode, where we adopt the VVC reference software, VTM version 18.0. Since VVC operates on the YUV color space by default, we first convert RGB images to YUV 4:4:4 color space using OpenCV.
Then, we execute the following command to compress the YUV image using VTM:
\begin{lstlisting}[language=Bash]
EncoderApp
    -c encoder_intra_vtm.cfg
    --InputFile={input file name}
    --BitstreamFile={bitstream file name}
    --SourceWidth={image width}
    --SourceHeight={image height}
    --InputChromaFormat=444
    --FrameRate=1
    --FramesToBeEncoded=1
    --QP={QP}
\end{lstlisting}
where QP is the quantization parameter, which we range from 15 to 50 with a step size of 1 to produce the PSNR-bpp curve.
We decode the bitstream using the following command:
\begin{lstlisting}[language=Bash]
DecoderApp
    --BitstreamFile={bitstream file name}
    --ReconFile={reconstruction file name}
    --OutputBitDepth=8
\end{lstlisting}
The reconstructed image has the YUV 4:4:4 format, and we converted it back to RGB space again using OpenCV.
The final PSNR is computed between the final RGB image and the original RGB image.

\begin{figure}[ht]
    \centering
    \subfloat[$\Lambda^{-1/3}$]{
        \includegraphics[width=0.47\linewidth]{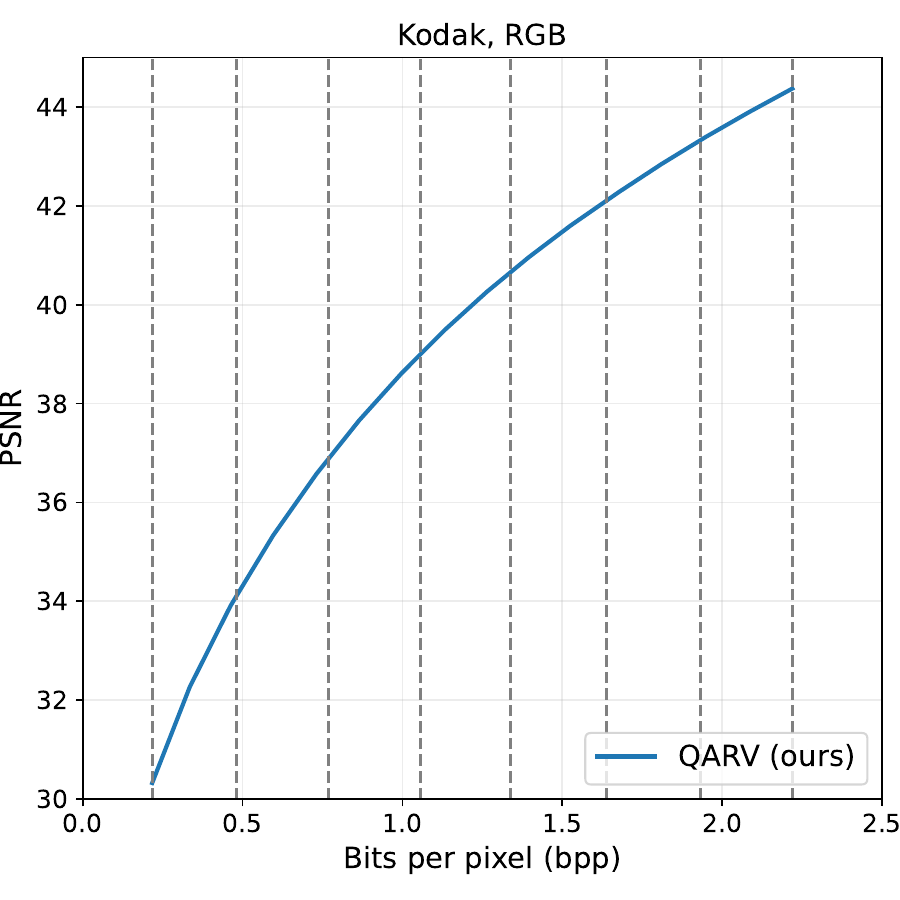}
        \label{fig:qarv_app_lmb_sample_cube}
    }
    \subfloat[$\log{\Lambda}$]{
        \includegraphics[width=0.47\linewidth]{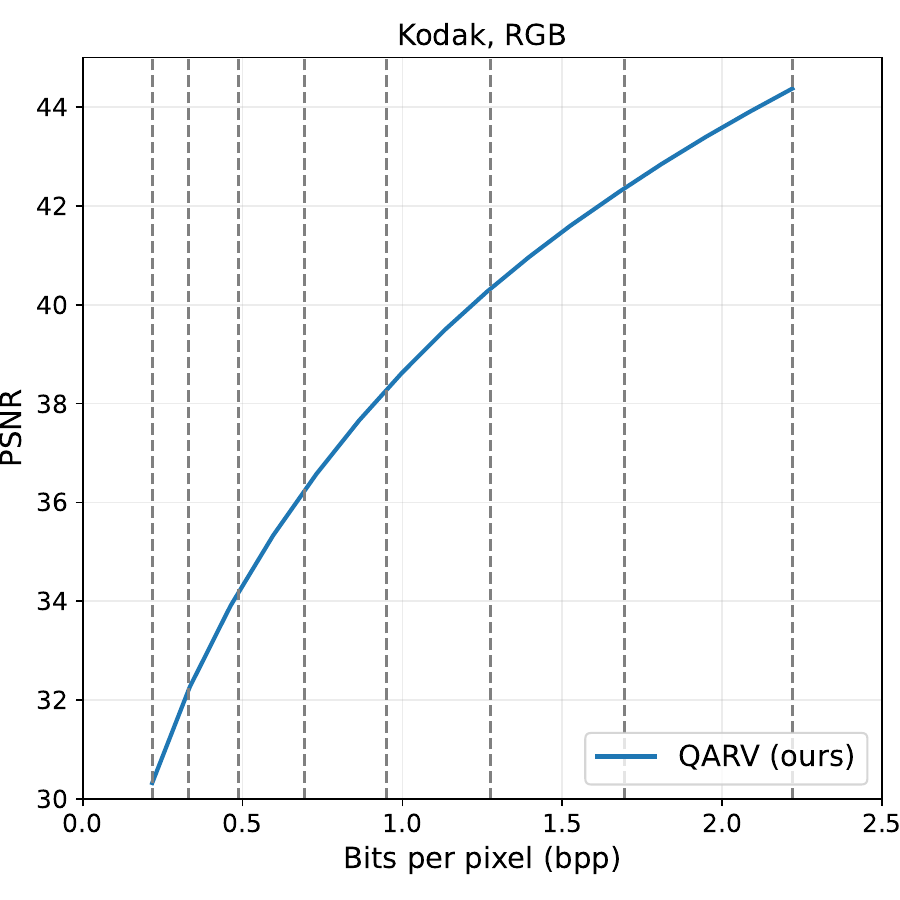}
        \label{fig:qarv_app_lmb_sample_log}
    }
    \caption{\textbf{Visualization of different choices of $p_\Lambda$}. We show equal-width bins in (a) the cube root space of $\Lambda$ and (b) the log space of $\Lambda$. We observe uniform sampling in the cube root space of $\Lambda$ results in a near-uniform sampling in the PSNR-bpp plane.}
    \label{fig:qarv_appendix_lmb_sample}
\end{figure}

\begin{figure*}[t]
    \centering
    \subfloat[]{
        \includegraphics[width=0.322\linewidth]{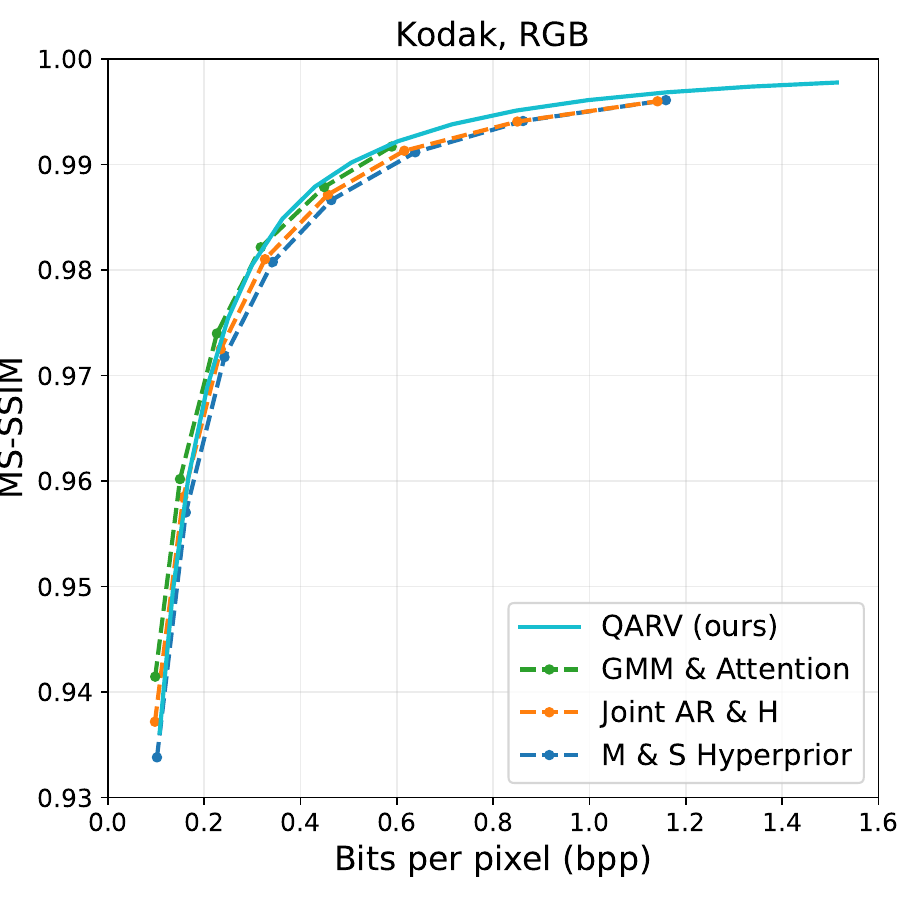}
    }
    \subfloat[]{
        \includegraphics[width=0.322\linewidth]{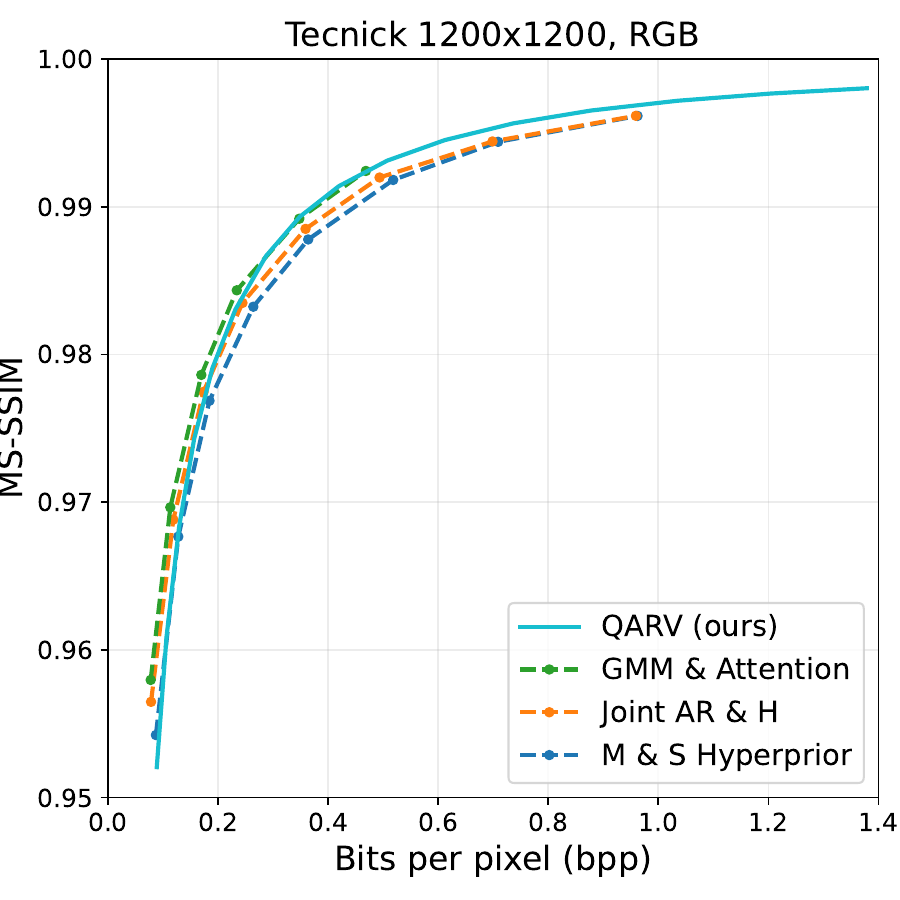}
    }
    \subfloat[]{
        \includegraphics[width=0.322\linewidth]{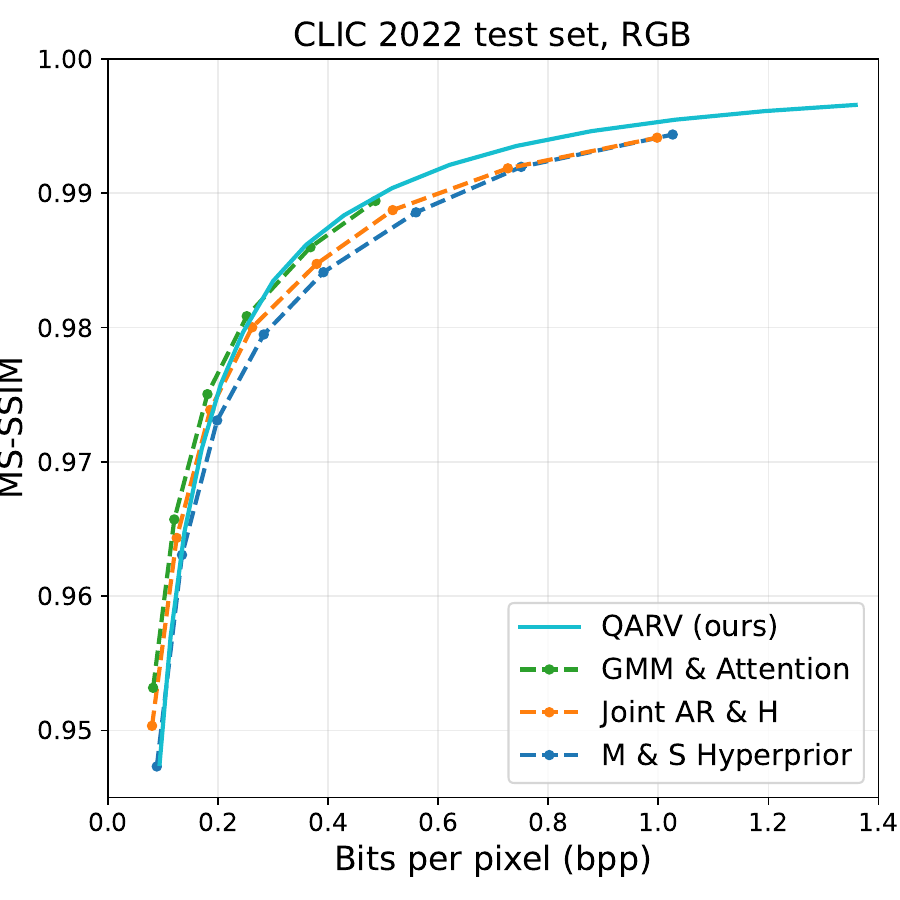}
    }
    \hfill
    \caption{
    \zhihao{
        \textbf{MS-SSIM vs. bpp results for three test sets:} (a) Kodak, (b) Tecnick, and (c) CLIC 2022 test split. Dashed lines are fixed-rate methods (\ie, separate models for different rates), and solid lines are variable-rate methods (\ie, a single model or program with adjustable rates).
        QARV is trained with reduced batch size and training iterations, but it still achieves comparable performance with previous baseline methods.
    }
    }
    \label{fig:qarv_app_ms_ssim}
\end{figure*}

\section{Variable-Rate Training}
\label{sec:qarv_appendix_var_rate_training}
In Sec.~\ref{sec:qarv_method2_rate_conditional}, we mentioned that we choose $p_\Lambda$ based on the observation that a uniform distribution in the cube root space of $\lambda$ leads to a near-uniform distribution in the rate space of the PSNR-bpp plane.
We evidence this in Fig.~\ref{fig:qarv_app_lmb_sample_cube}, where we show the PSNR-bpp curve produced by QARV along with edges of equal-mass bins of $p_\Lambda$.
Specifically, we obtain the bin edges by first computing $\lambda_i, \forall i=0, 1, ..., M$:
\begin{equation}
\begin{aligned}
    \delta_i &= \frac{i}{M} \cdot \lambda_\text{low}^{-1/3} + \frac{M - i}{M} \cdot \lambda_\text{high}^{-1/3}
    \\
    \lambda_i &= \delta_i^3.
\end{aligned}
\end{equation}
We then evaluate QARV at each $\lambda_i, \forall i=0, 1, ..., M$, where we set $M=7$ for visualization.
The resulting rates are shown as vertical, black dashed lines in Fig.~\ref{fig:qarv_app_lmb_sample_cube}.
In other words, the bins in Fig.~\ref{fig:qarv_app_lmb_sample_cube} have an equal probability mass under $p_\Lambda$.
As a result, sampling from $p_\Lambda$ leads to a near-uniform sampling in the rate space, and thus training covers all rate range with approximately the same weight.

As a comparison, Fig.~\ref{fig:qarv_app_lmb_sample_log} show the equal-mass bin edges of the pdf that is uniform in its log space, \ie,
\begin{equation}
\begin{aligned}
    \delta_i &= \frac{i}{M} \cdot \log \lambda_\text{low} + \frac{M - i}{M} \cdot \log \lambda_\text{high}
    \\
    \lambda_i &= \exp{\delta_i}.
\end{aligned}
\end{equation}
We can observe from Fig.~\ref{fig:qarv_app_lmb_sample_log} that this pdf has a clear bias towards lower rates.
That is, the lower-rate region is sampled much more frequently than higher-rate regions.
In Table~\ref{table:qarv_app_lmb_sampling}, we show the training results of this sampling strategy (second row) compared to our $p_\Lambda$ (first row), where we train both methods for 1M iterations, and the BD-rate is w.r.t. VTM 18.0.
Results show that our $p_\Lambda$ performs better in BD-rate.
We conjecture that lower-rate compression is intrinsically a simpler problem than high-rate compression (since high-rate compression involves learning high-frequency pixel details), and thus one should not assign higher probability density to sample lower rates during training.

\begin{table}[t]
\centering
\caption{Comparison of the Choices of $p_\Lambda$.}
\vspace{-8pt}
\begin{tabular}{c|c}
\hline
\textbf{Choice of $p_\Lambda$} & \textbf{Kodak BD-rate (\%)} \\ \hline
$\Lambda = U(\lambda_\text{low}^{-1/3}, \lambda_\text{high}^{-1/3})^3$ & \textbf{-4.74}                     \\
$\Lambda = \exp{U(\log \lambda_\text{low}, \log \lambda_\text{high})}$ & -4.28                     \\ \hline
\end{tabular}
\label{table:qarv_app_lmb_sampling}
\end{table}



\def\dmsssim{d_\text{M-S}}

\zhihao{
\section{Training Loss Function for MS-SSIM}
The choice of the distortion metric in our framework is arbitrary.
For example, one could use MS-SSIM~\cite{wang2004msssim} instead of PSNR as the evaluation metric.
To optimize QARV for MS-SSIM during training, we use the training loss function defined in Eq.~\eqref{eq:qarv_method_loss_var_rate} with the following distortion function based on MS-SSIM:
\begin{equation}
    \dmsssim(x, \hat{x}) = 1 - \text{MS-SSIM}(x, \hat{x}).
\end{equation}
Since MS-SSIM takes a value between 0 and 1 (1 means perfect quality), we have $\dmsssim(x, \hat{x}) \in [0, 1]$ with $\dmsssim(x, \hat{x}) = 0$ if and only if $x = \hat{x}$.

\textbf{Experiment:} We use the MS-SSIM implementation from an open-source software repository\footnote{https://github.com/VainF/pytorch-msssim}.
We train our QARV model with the MS-SSIM loss, while all other settings are the same as the ablation study experiments (the right column in Table~\ref{table:qarv_appendix_hyp_param}).
Results are shown in Fig.~\ref{fig:qarv_app_ms_ssim}.
We observe that QARV achieves slightly lower MS-SSIM with the baseline methods at low rates (bpp $< 0.5$) and clearly higher MS-SSIM at high rates (bpp $> 0.5$).
This observation is consistent with the PSNR results reported in Sec.~\ref{sec:qarv_exp_sota_comparison}, which we attribute to the fact that QARV employs high-dimensional latent variables and thus naturally biases towards high rates.

Overall, we can see that QARV easily adapts to the MS-SSIM metric.
Note that we did not tune the model or training hyperparameters for MS-SSIM, and the model for MS-SSIM is trained with the reduced training recipe.
With hyperparameter tuning and a longer training time, we believe QARV can be adapted to various distortion metrics with performance competitive to existing best methods.
}

\begin{figure}[t]
    \centering
    \subfloat[Training loss vs. iterations]{
        \includegraphics[width=0.47\linewidth]{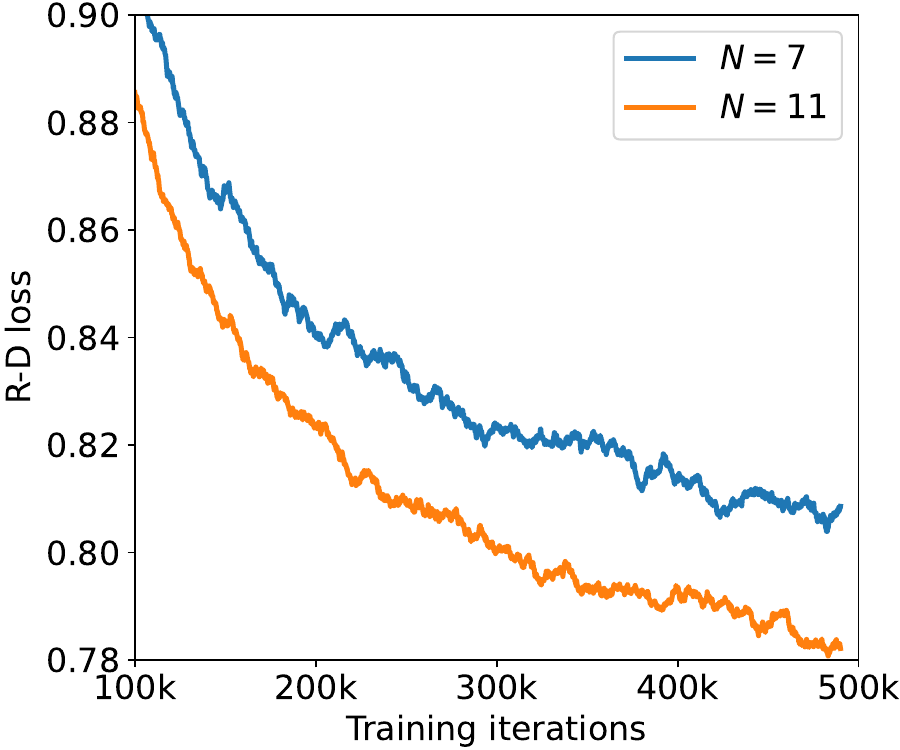}
        \label{fig:qarv_app_n11_train}
    }
    \subfloat[BD-rate vs. iterations]{
        \includegraphics[width=0.47\linewidth]{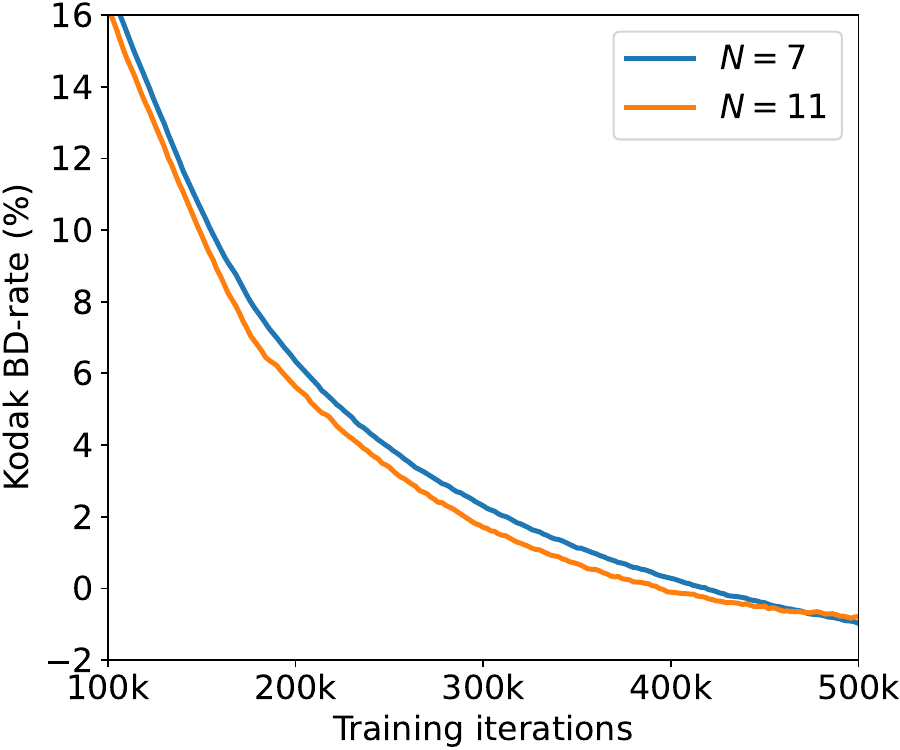}
        \label{fig:qarv_app_n11_test}
    }
    \caption{
        \zhihao{We observe that the training loss (\ie, rate-distortion) and the evaluation BD-rate are not always consistent. For example, a model with $N=11$ latent variables achieves a lower training loss than one with $N = 7$, but it is not true for BD-rate. BD-rate is w.r.t. VTM 18.0.}
    }
    \label{fig:qarv_app_n11_traintest}
\end{figure}

\zhihao{
\section{Number of Latent Variables}
\label{sec:qarv_appendix_num_latent}
Recall that QARV employs a total number of $N$ latent variables (where $N$ can be set arbitrarily) with various resolutions.
In our main model, we choose $N = 9$ and assign $[1, 2, 3, 3]$ of them to the features that are $64\times, 32\times, 16\times, 8\times$ downsampled w.r.t. the original image, respectively.
Please see Fig.~\ref{fig:qarv_method_architecture} for the illustration.

To study the impact of $N$, we train models with $N \in \{ 3, 5, 7, 9, 11 \}$ and report results in Table~\ref{table:qarv_app_z_number}.
Note that the ``distribution of variables'' in the table refers to the number of latent variables that are $64\times, 32\times, 16\times, 8\times$ downsampled w.r.t. the original image, respectively.
The first observation is that as $N$ increases from $3$ to $9$, more latent variables achieve better BD-rate at the expense of more model parameters.
This is expected because each latent variable block contains multiple neural network modules, which improve model expressiveness but increases the model size.

However, as we see from the last row of Table~\ref{table:qarv_app_z_number}, increasing $N$ from 9 to 11 does not give a better BD-rate.
We observe that this is caused by an issue similar to overfitting, where training loss is inconsistent with evaluation metrics.
We show the training loss (on COCO) and evaluation BD-rate (on Kodak) as functions of training iterations in Fig.~\ref{fig:qarv_app_n11_train} and Fig.~\ref{fig:qarv_app_n11_test}, respectively.
When comparing the case of $N = 11$ with $N = 7$, we see that $N = 11$ is clearly better than $N = 7$ in terms of training loss at all training iterations.
However, this is not true for the BD-rate at the 500k training iteration.
As the training loss is not equivalent to the test-time BD-rate, it is possible that the model learns to minimize the training loss without improving BD-rate, and future research is needed to address this issue.
}

\begin{table}[t]
\centering
\caption{\zhihao{Impact of the Number of Latent Variables.}}
\vspace{-8pt}
\begin{tabular}{cc|c|c}
\hline
$N$ & Distribution of variables & \textbf{Params.} & \textbf{Kodak BD-rate (\%)} \\ \hline
3   & $[1,0,1,1]$   & 61.0M   & 8.030  \\
5   & $[1,1,1,2]$   & 71.9M   & 1.140  \\
7   & $[1,2,2,2]$   & 85.6M   & -0.716  \\
9   & $[1,2,3,3]$   & 93.4M   & -0.433  \\
11  & $[1,2,4,4]$   & 101.3M  & -0.513  \\ \hline
\end{tabular}
\label{table:qarv_app_z_number}
\end{table}

\section{Rate Distribution over Latent Variables}
\begin{figure*}[t]
    \centering
    \subfloat[Bpp consumption of each latent variable under various $\lambda$.]{
        \includegraphics[width=0.84\linewidth]{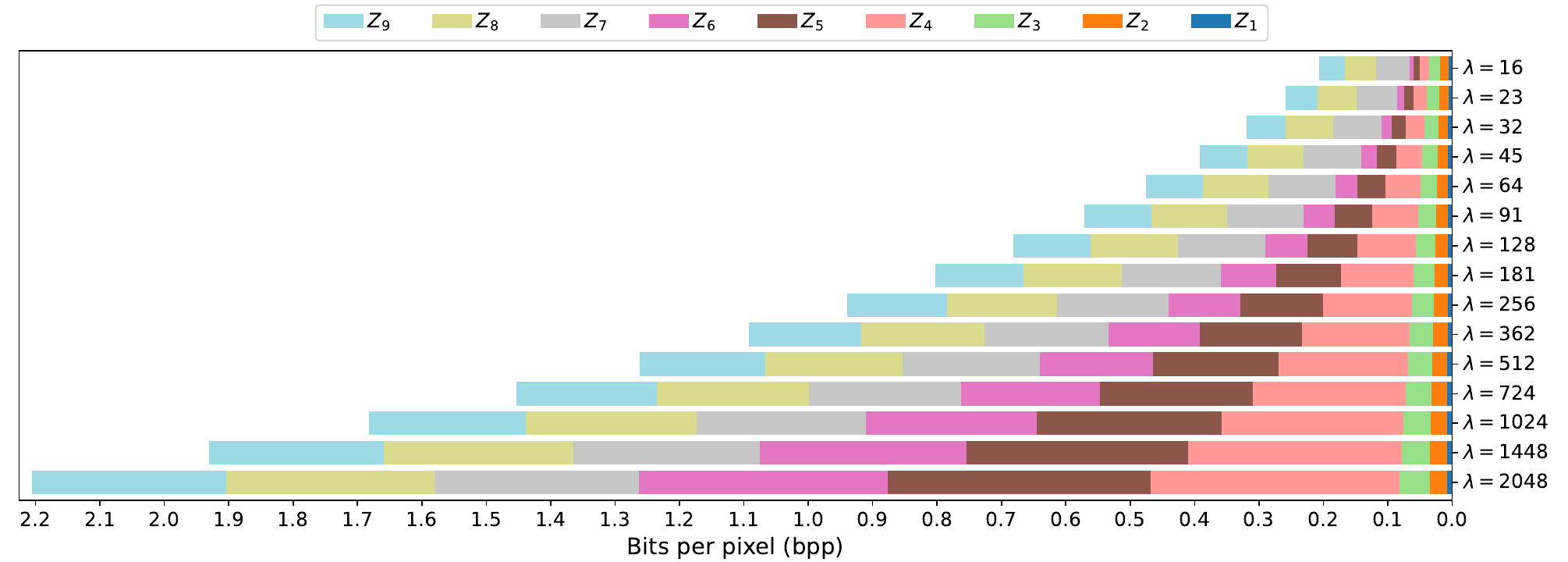}
        \label{fig:qarv_app_bpp_distribution_abs}
    }
    \hfill
    \subfloat[Rate percentage of each latent variable under various $\lambda$.]{
        \includegraphics[width=0.84\linewidth]{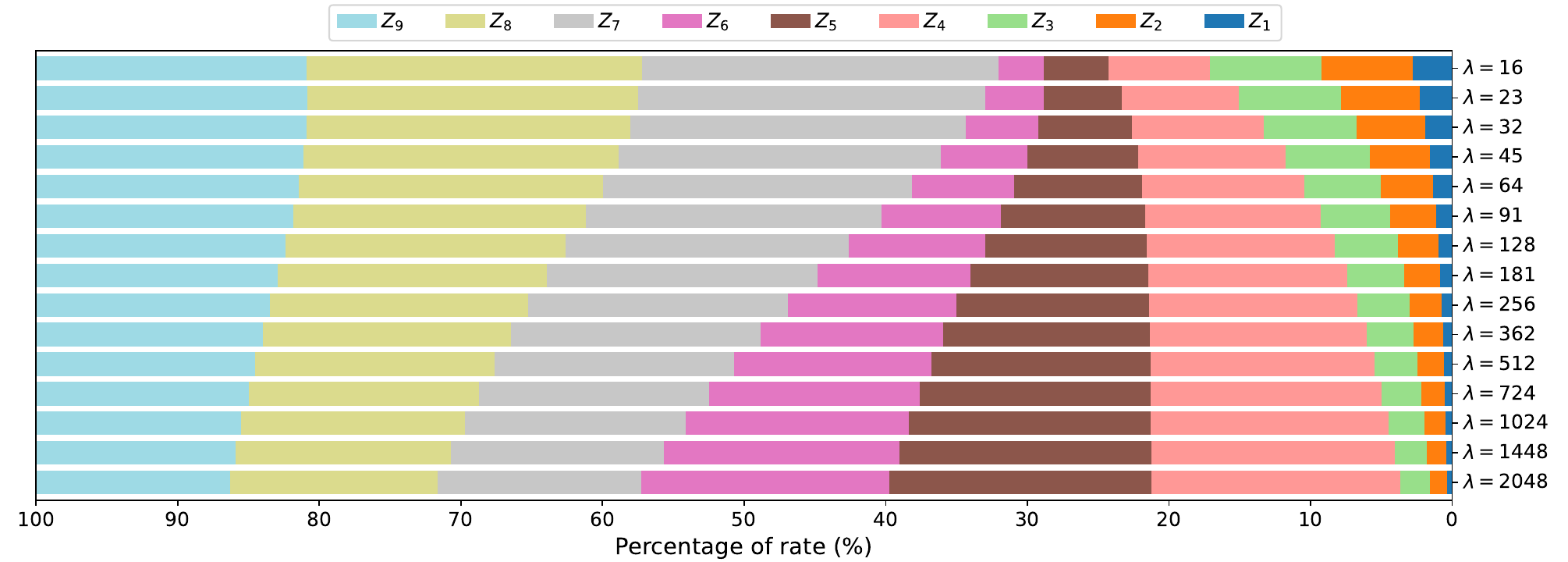}
        \label{fig:qarv_app_bpp_distribution_norm}
    }
    \caption{\textbf{Rate distribution over latent variables $Z_{1:N}$ under various $\lambda$.} The x-axis is bpp in (a) and percentage in (b).  In both figures, rows correspond to different $\lambda$ values. In each row, different color represents different latent variables: $Z_1$ has the smallest spatial dimension ($64\times$ downsampled w.r.t. the input image), and $Z_9$ has the largest spatial dimension ($8\times$ downsampled w.r.t. the input image). }
    \label{fig:qarv_app_bpp_distribution}
\end{figure*}

Recall that the QARV model produces $N=9$ bitstreams for each image.
Each bitstream corresponds to a different latent variable, and the lengths of the bitstreams vary depending on the input $\lambda$.
To better understand how the rate distributes over latent variables and how $\lambda$ impacts such distributions, we visualize them in Fig.~\ref{fig:qarv_app_bpp_distribution}, where the results are averaged on all Kodak images.

Several interesting facts can be observed from Fig.~\ref{fig:qarv_app_bpp_distribution_abs}.
First, we see that a larger $\lambda$ leads to a higher overall rate.
When looking at the rate for each individual latent variable, one can observe that the rates for variables $Z_{1:3}$ stay approximately unchanged as $\lambda$ ranges from 16 to 2048.
However, the rates for $Z_{4:9}$ clearly increase monotonically as $\lambda$ increases from 16 to 2048, and for high $\lambda$, the overall rate is dominated by higher-dimensional latent variables.
We also depict in Fig.~\ref{fig:qarv_app_bpp_distribution_norm} the percentage (instead of absolute values) of the rate each latent variable consumes w.r.t. the overall rate of an image.
The observation is consistent with Fig.~\ref{fig:qarv_app_bpp_distribution_abs}. The majority of the bit rate is produced by high-dimensional latent variables, regardless of the input $\lambda$.

\begin{figure}[t]
    \centering
    \includegraphics[width=0.8\linewidth]{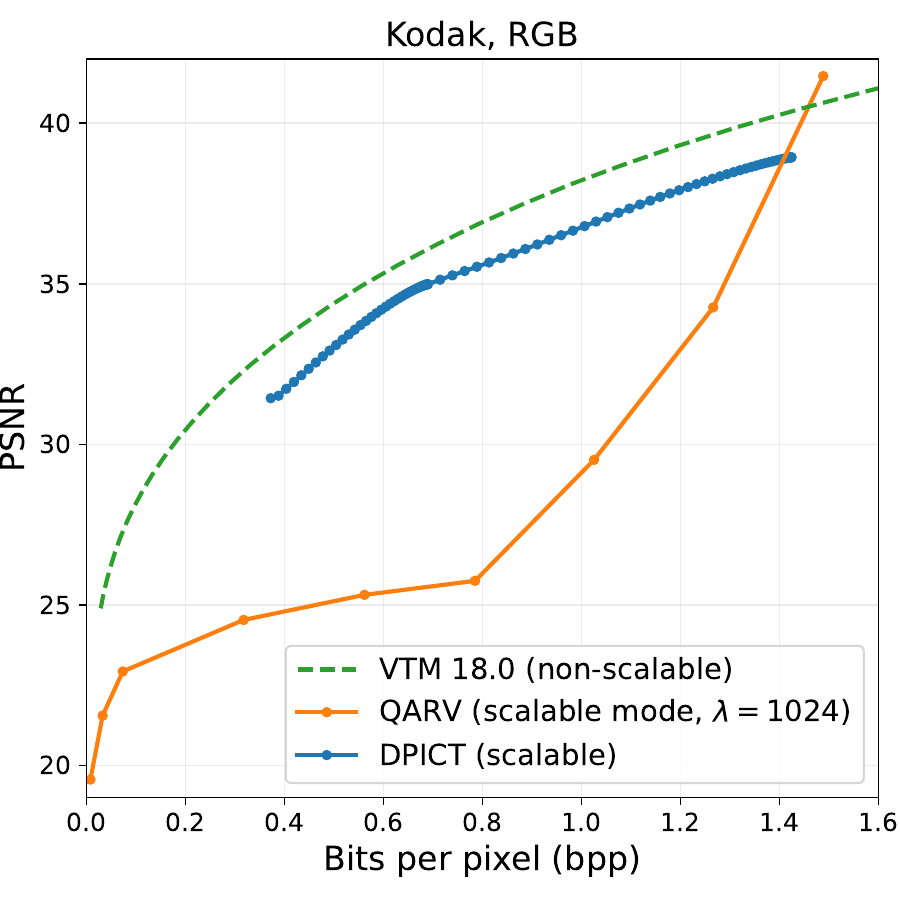}
    \caption{
        \zhihao{QARV supports scalable coding, although it is not optimized for it. Solid lines are scalable coding methods, and dashed lines are non-scalable methods (showing for reference).}
    }
    \label{fig:qarv_app_scalable}
\end{figure}

\zhihao{
\section{Scalable Coding using QARV}
Recall that QARV encodes every input image $X$ into $N$ latent variables $Z_{1:N}$.
To achieve scalable coding, we encode the image into the first subset of $i$ latent variables ($Z_{\leq i}$) and ignore the rest ($Z_{>i}$).
Then, $Z_{\leq i}$ is sent over the communication channel, and on the decoding side, we decode $Z_{\leq i}$ and use the prior mean for $Z_{>i}$.
As soon as the decoder receives $Z_{i+1}$, the decoder refines the reconstruction by $Z_{i+1}$, and this process effectively implements scalable coding.

We report the scalable coding performance of our main QARV model in Fig.~\ref{fig:qarv_app_scalable}. 
We observe that QARV is worse than recent scalable image coding methods such as DPICT~\cite{lee2022dpict}, which is expected since QARV is not optimized for scalable R-D performance.
Scalable coding is merely an optional feature rather than an intended use case for QARV.
That said, there exists the possibility to optimize the QARV architecture for scalable coding, \eg, by imposing an R-D loss function for every latent variable separately, but it is out of the scope of this paper, and we leave it to future research.
}

\end{document}